\global\def\draftcontrol{0}

   \def\versionno{Gauged $\caln=2$ and GKG}

\catcode`\@=11

\expandafter\ifx\csname draftcontrol\endcsname\relax\global\def\draftcontrol{0}
\fi

{\count255=\time\divide\count255 by 60
\xdef\hourmin{\number\count255}
\multiply\count255 by-60\advance\count255 by\time
\xdef\hourmin{\hourmin:\ifnum\count255<10 0\fi\the\count255}}
\def\draftdate{\number\month/\number\day/\number\year\ \ \ \hourmin }

\newcommand\makepapertitle{\par
  \begingroup
    \renewcommand\thefootnote{\@fnsymbol\c@footnote}%
    \def\@makefnmark{\rlap{\@textsuperscript{\normalfont\@thefnmark}}}%
    \long\def\@makefntext##1{\parindent 1em\noindent
            \hb@xt@1.8em{%
                \hss\@textsuperscript{\normalfont\@thefnmark}}##1}%
     \newpage
     \global\@topnum\z@   
     \@makepapertitle
     \thispagestyle{empty}\@thanks
  \endgroup
  \setcounter{footnote}{0}%
  \global\let\thanks\relax
  \global\let\makepapertitle\relax
  \global\let\@makepapertitle\relax
  \global\let\@thanks\@empty
  \global\let\@author\@empty
  \global\let\@date\@empty
  \global\let\@title\@empty
  \global\let\title\relax
  \global\let\author\relax
  \global\let\date\relax
  \global\let\and\relax
  \def\version{\let\version\@version\@gobble}
}
\def\@makepapertitle{%
  \newpage
   \ifnum\draftcontrol=1 {}
   \version\versionno
   \vskip 3em%
   \else
   \hfill\hbox to 3cm {\parbox{4cm}{\@pubnum}\hss}%
   \vskip 3em%
   \fi
   \begin{center}%
   \let \footnote \thanks
     {\LARGE {\@title}}%
     \vskip 1.5em%
     {\normalsize
       \lineskip .5em%
       \begin{tabular}[t]{c}%
         \@author
       \end{tabular}\par}%
     \vskip 1.5em%
     {\@bstract}%
     \end{center}%
     \vskip 1.5em
     \@date%
   \par
}

\gdef\@pubnum{}
\def\pubnum#1{%
  \gdef\@pubnum{#1}}

\gdef\@bstract{}
\def\Abstract#1{%
  \gdef\@bstract{%
   \parbox{\textwidth-0pc}{%
   \centerline{\bf Abstract}\penalty1000%
\noindent
\renewcommand\baselinestretch{1.0}%
{#1}}}
}

\def\ps@paper{\let\@mkboth\@gobbletwo%
     \ifnum\draftcontrol=1
        \def\@oddfoot{\hbox to \textwidth{\tiny \versionno \hfil\tiny\draftdate}%
        \hskip -\textwidth \hbox to \textwidth{\hfil\rm\thepage\hfil}}%
     \else\def\@oddfoot{\hbox to \textwidth{\hfil\rm\thepage\hfil}}
     \fi
     \let\@evenfoot\@oddfoot
}

\def\@version#1{\ifnum\draftcontrol=1
\typeout{}\typeout{#1}\typeout{}
\vskip3mm\centerline{\hbox{\fbox{\normalsize{\tt DRAFT -- #1 -- }
                   {\draftdate}}}}\vskip3mm
\fi}
\let\version\@version
\long\def\eqlabel#1{\ifnum\draftcontrol=1
                    \tag@false  
                    \tag*{(\theequation) \hbox to -0.2cm{\hspace{0cm}\small{#1}\hss}}
                    \refstepcounter{equation}
                    \edef\@currentlabel{\theequation}
                    \ltx@label{#1}          
                    \else
                    \label{#1}
                    \fi
                    }
\let\st@bibitem\@bibitem
\let\st@lbibitem\@lbibitem
\ifnum\draftcontrol=1
  \def\@bibitem#1{%
    \st@bibitem{#1}\a@@label{#1}\ignorespaces}
  \def\@lbibitem[#1]#2{%
    \st@lbibitem[#1]{#2}\a@@label{#2}\ignorespaces}
  \def\a@@label#1{%
    \gdef\a@lab{\smash{\normalfont\small#1}}
    \ifvmode
      \if@inlabel
        \global\setbox\@labels\hbox{%
          \llap{\a@lab\let\a@lab\relax
                \kern\@totalleftmargin\kern\marginparsep}%
          \box\@labels}%
      \fi
    \fi}
\fi

\documentclass[12pt,letterpaper]{article}
\usepackage{amsmath,amssymb,array,calc,rotating,epsfig,psfrag}
\usepackage[nosort]{cite}

\ifnum\draftcontrol=1
\fi

\renewcommand\baselinestretch{1.25}
\renewcommand\section{\@startsection {section}{1}{\z@}%
                                   {-3.5ex \@plus -1ex \@minus -.2ex}%
                                   {2.3ex \@plus.2ex}%
                                   {\normalfont\large\bfseries}}
\renewcommand\subsection{\@startsection{subsection}{2}{\z@}%
                                   {-3.25ex\@plus -1ex \@minus -.2ex}%
                                   {1.5ex \@plus .2ex}%
                                   {\normalfont\normalsize\bfseries}}
\renewcommand\subsubsection{\@startsection{subsubsection}{3}{\z@}%
                                   {-3.25ex\@plus -1ex \@minus -.2ex}%
                                   {1.5ex \@plus .2ex}%
                                   {\normalfont\normalsize\it}}
\renewcommand\paragraph{\@startsection{paragraph}{4}{\z@}%
                                   {-3.25ex\@plus -1ex \@minus -.2ex}%
                                   {1.5ex \@plus .2ex}%
                                   {\normalfont\normalsize\bf}}

\def\revise#1       {\raisebox{-0em}{\rule{3pt}{1em}}%
                     \marginpar{\raisebox{.5em}{\vrule width3pt\
                     \vrule width0pt height 0pt depth0.5em
                     \hbox to 0cm{\hspace{0cm}{%

\parbox[t]{4em}{\raggedright\footnotesize{#1}}}\hss}}}}

\def\caln         {{\cal N}}

\def\del          {\partial}

\def\tr           {\mathop{\rm Tr}}

\def\half{{\frac12}}

\def\sqr#1#2{{\vcenter{\vbox{\hrule height.#2pt
 \hbox{\vrule width.#2pt height#1pt \kern#1pt
 \vrule width.#2pt}\hrule height.#2pt}}}}

\def\a{\alpha}
\def\b{\beta}
\def\r{\rho}

\def\x{\xi}

\def\m{\mu}
\def\g{\gamma}

\def\n{\nu}
\def\bn{\bar{\nu}}
\def\bm{\bar{\mu}}

\def\pp{{\mathchoice
          {
              \kern 1pt%
              \raise 1pt
              \vbox{\hrule width5pt height0.4pt depth0pt
                    \kern -2pt
                    \hbox{\kern 2.3pt
                          \vrule width0.4pt height6pt depth0pt
                          }
                    \kern -2pt
                    \hrule width5pt height0.4pt depth0pt}%
                    \kern 1pt
           }
            {
              \kern 1pt%
              \raise 1pt
              \vbox{\hrule width4.3pt height0.4pt depth0pt
                    \kern -1.8pt
                    \hbox{\kern 1.95pt
                          \vrule width0.4pt height5.4pt depth0pt
                          }
                    \kern -1.8pt
                    \hrule width4.3pt height0.4pt depth0pt}%
                    \kern 1pt
            }
            {
              \kern 0.5pt%
              \raise 1pt
              \vbox{\hrule width4.0pt height0.3pt depth0pt
                    \kern -1.9pt  
                    \hbox{\kern 1.85pt
 \vrule width0.3pt height5.7pt depth0pt
                          }
                    \kern -1.9pt
                    \hrule width4.0pt height0.3pt depth0pt}%
                    \kern 0.5pt
            }
           {
              \kern 0.5pt%
              \raise 1pt
              \vbox{\hrule width3.6pt height0.3pt depth0pt
                    \kern -1.5pt
                    \hbox{\kern 1.65pt
                          \vrule width0.3pt height4.5pt depth0pt
                          }
                    \kern -1.5pt
                    \hrule width3.6pt height0.3pt depth0pt}%
                    \kern 0.5pt
            }
        }}

 \def\mm{{\mathchoice
                       {
                             \kern 1pt
               \raise 1pt    \vbox{\hrule width5pt height0.4pt depth0pt
                                  \kern 2pt
                                  \hrule width5pt height0.4pt depth0pt}
                             \kern 1pt}
                       {
                            \kern 1pt
               \raise 1pt \vbox{\hrule width4.3pt height0.4pt depth0pt
                                  \kern 1.8pt
                                  \hrule width4.3pt height0.4pt depth0pt}
                             \kern 1pt}
                       {
                            \kern 0.5pt
               \raise 1pt
                            \vbox{\hrule width4.0pt height0.3pt depth0pt
                                  \kern 1.9pt
                                  \hrule width4.0pt height0.3pt depth0pt}
                            \kern 1pt}
                       {
                           \kern 0.5pt
             \raise 1pt  \vbox{\hrule width3.6pt height0.3pt depth0pt
                                  \kern 1.5pt
                                  \hrule width3.6pt height0.3pt depth0pt}
                           \kern 0.5pt}
}}
 \def\ad{{\kern0.5pt
                  \alpha \kern-5.05pt \raise5.8pt\hbox{$\textstyle.$}\kern
 0.5pt}}
 \def\bd{{\kern0.5pt
                   \beta \kern-5.05pt \raise5.8pt\hbox{$\textstyle.$}\kern
 0.5pt}}

 \def\qd{{\kern0.5pt
                   q \kern-5.05pt \raise5.8pt\hbox{$\textstyle.$}\kern
 0.5pt}}
 \def\Dot#1{{\kern0.5pt
     {#1} \kern-5.05pt \raise5.8pt\hbox{$\textstyle.$}\kern
 0.5pt}}

\catcode`\@=12

\begin{document}


\topmargin=-0.50in

\newcommand{\be}{\begin{equation}}
\newcommand{\ee}{\end{equation}}
\newcommand{\beq}{\begin{equation}}
\newcommand{\eeq}{\end{equation}}
\newcommand{\ba}{\begin{eqnarray}}
\newcommand{\ea}{\end{eqnarray}}
\newcommand{\nn}{\nonumber}

\def\vol{\bf vol}
\def\Vol{\bf Vol}
\def\del{{\partial}}
\def\vev#1{\left\langle #1 \right\rangle}
\def\cn{{\cal N}}
\def\co{{\cal O}}
\def\IC{{\mathbb C}}
\def\IR{{\mathbb R}}
\def\IZ{{\mathbb Z}}
\def\RP{{\bf RP}}
\def\CP{{\bf CP}}
\def\Poincare{{Poincar\'e }}
\def\tr{{\rm tr}}
\def\tp{{\tilde \Phi}}
\def\Y{{\bf Y}}
\def\te{\theta}
\def\bX{\bf{X}}

\def\TL{\hfil$\displaystyle{##}$}
\def\TR{$\displaystyle{{}##}$\hfil}
\def\TC{\hfil$\displaystyle{##}$\hfil}
\def\TT{\hbox{##}}
\def\HLINE{\noalign{\vskip1\jot}\hline\noalign{\vskip1\jot}} 
\def\seqalign#1#2{\vcenter{\openup1\jot
  \halign{\strut #1\cr #2 \cr}}}
\def\lbldef#1#2{\expandafter\gdef\csname #1\endcsname {#2}}
\def\eqn#1#2{\lbldef{#1}{(\ref{#1})}%
\begin{equation} #2 \label{#1} \end{equation}}
\def\eqalign#1{\vcenter{\openup1\jot
    \halign{\strut\span\TL & \span\TR\cr #1 \cr
   }}}
\def\eno#1{(\ref{#1})}
\def\href#1#2{#2}
\def\half{{1 \over 2}}

\def\ads{{\it AdS}}
\def\adsp{{\it AdS}$_{p+2}$}
\def\cft{{\it CFT}}

\newcommand{\ber}{\begin{eqnarray}}
\newcommand{\eer}{\end{eqnarray}}

\newcommand{\bea}{\begin{eqnarray}}
\newcommand{\eea}{\end{eqnarray}}

\newcommand{\beqar}{\begin{eqnarray}}
\newcommand{\cN}{{\cal N}}
\newcommand{\cO}{{\cal O}}
\newcommand{\cA}{{\cal A}}
\newcommand{\cT}{{\cal T}}
\newcommand{\cF}{{\cal F}}
\newcommand{\cC}{{\cal C}}
\newcommand{\cR}{{\cal R}}
\newcommand{\cW}{{\cal W}}
\newcommand{\eeqar}{\end{eqnarray}}
\newcommand{\lm}{\lambda}\newcommand{\Lm}{\Lambda}
\newcommand{\eps}{\epsilon}


\newcommand{\nonu}{\nonumber}
\newcommand{\oh}{\displaystyle{\frac{1}{2}}}
\newcommand{\dsl}
  {\kern.06em\hbox{\raise.15ex\hbox{$/$}\kern-.56em\hbox{$\partial$}}}
\newcommand{\as}{\not\!\! A}
\newcommand{\ps}{\not\! p}
\newcommand{\ks}{\not\! k}
\newcommand{\D}{{\cal{D}}}
\newcommand{\dv}{d^2x}
\newcommand{\Z}{{\cal Z}}
\newcommand{\N}{{\cal N}}
\newcommand{\Dsl}{\not\!\! D}
\newcommand{\Bsl}{\not\!\! B}
\newcommand{\Psl}{\not\!\! P}
\newcommand{\eeqarr}{\end{eqnarray}}
\newcommand{\ZZ}{{\rm \kern 0.275em Z \kern -0.92em Z}\;}

\def\s{\sigma}
\def\a{\alpha}
\def\b{\beta}
\def\r{\rho}
\def\d{\delta}
\def\g{\gamma}
\def\G{\Gamma}
\def\ep{\epsilon}
\def\P{\Phi}
\def\o{\omega}
\def\e{\epsilon}
\def\q{\theta}

\makeatletter \@addtoreset{equation}{section} \makeatother
\renewcommand{\theequation}{\thesection.\arabic{equation}}

\def\be{\begin{equation}}
\def\ee{\end{equation}}
\def\bea{\begin{eqnarray}}
\def\eea{\end{eqnarray}}
\def\m{\mu}
\def\n{\nu}
\def\g{\gamma}
\def\p{\phi}
\def\L{\Lambda}
\def \W{{\cal W}}
\def\bn{\bar{\nu}}
\def\bm{\bar{\mu}}
\def\bw{\bar{w}}
\def\ba{\bar{\alpha}}
\def\bb{\bar{\beta}}
\def\bd{\bar{D}}
\def\bP{\bar{\Phi}}
\def\bL{\bar{\Lambda}}
\def\ra{\rightarrow}
\def\Ra{\Rightarrow}
\def\mcal{\mathcal}

\begin{titlepage}

\version\versionno

\leftline{\tt hep-th/0610116}

\vskip -.8cm

\rightline{\small{\tt MCTP-06-21}}
\rightline{\small{\tt UMDEPP 06-050}}
\rightline{\small{\tt NSF-KITP-06-70}}

\vskip 1.7 cm

\centerline{\bf \Large Gauged $(2,2)$ Sigma Models}

\vspace{.5cm}

\centerline{\bf \Large and Generalized K\"ahler Geometry}

\vskip .2cm
\vskip 1cm
{\large }
\vskip 1cm

\centerline{\large  Willie Merrell${}^1$, Leopoldo A. Pando Zayas${}^2$
and Diana Vaman${}^2$  }

\vskip .5cm
\centerline{\it ${}^1$ Department of Physics}
\centerline{ \it University of Maryland}
\centerline{\it College Park, MD 20472}

\vskip .2cm
\centerline{\it ${}^2$ Michigan Center for Theoretical
Physics}
\centerline{ \it Randall Laboratory of Physics, The University of
Michigan}
\centerline{\it Ann Arbor, MI 48109}

\vspace{1cm}

\begin{abstract}
We gauge the  $(2,2)$ supersymmetric non-linear sigma model
whose target space has bihermitian structure $(g, B, J_{\pm})$ with noncommuting complex
structures. The bihermitian geometry
is realized by a sigma model which is written in terms of
$(2,2)$ semi-chiral superfields.
We discuss the moment map, from the perspective of the gauged sigma model
action and from the integrability condition for a Hamiltonian vector
field. We show that for a concrete example, the $SU(2)\times U(1)$ WZNW model,
as well as for the sigma models with almost product structure, the moment map
can be used together with the corresponding Killing vector to form an
element of
$T\oplus T^*$ which lies in the eigenbundle of the generalized almost
complex structure.
Lastly, we discuss T-duality at the level of a $(2,2)$ sigma model
involving semi-chiral superfields and present an explicit example.
\end{abstract}



\end{titlepage}
\section{Introduction}
\indent
{\it Superfield representations and geometry:} The connection between
geometry and supersymmetry has been an evolving and recurring theme since
the early days of supersymmetry.  The earliest example was the connection
between K\"ahler geometry and supersymmetric non linear sigma models
introduced in a seminal paper by Zumino \cite{zumino} who considered the
conditions for existence of ${\cal N}=(2,2)$ on two-dimensional nonlinear
sigma models.
This notion
was further developed by \'{A}lvarez-Gaum\'{e} and Freedman who showed that
further extensions of supersymmetry to ${\cal N}=(4,4)$ required the sigma
model metric to be hyperk\"{a}hler \cite{ag-f}.

Other important geometric structures were understood in this context
including the moment map as well as symplectic and K\"ahler reductions. Many
of these structures were devoloped independently in the mathematical and physics
literature. The use of a Legendre transform and a symplectic quotient in the
study of hyperk\"{a}hler geometry arose from their use in supersymmetric
sigma models \cite{lr,rt,cf}. In the context of hyperk\"ahler geometry a
comprehensive review was presented in \cite{hypereview}.

An important step in understanding the general  structure behind geometry
and supersymmetry was taken in \cite{ag-f-2} which presented a classification
of the geometries consistent with extended supersymmetry paying particular
attention to the type of superfield representations involved. Perhaps the
clearest example of the connection between geometry and the superfield
representations arose out of the study of two-dimensional $\mcal{N}=(2,2)$
supersymmetric models discovered by Gates, Hull, and Ro\v{c}ek \cite{ghr}.
This work introduced twisted chiral superfields on the supersymmetry side
and bihermitian geometry on the mathematical side. Subsequently, Buscher,
Lindstr\"{o}m, and Ro\v{c}ek \cite{blr} expanded in this direction.  These
results showed that the underlying element in the relation between the
amount of supersymmetry and different versions of complex geometry is to a
large extent determined by the superfield representations involved.

{\it Generalized complex geometry and $(2,2)$ supersymmetry:} What would
complex geometry look like if instead of considering structures associated
with the tangent bundle $T$ one considers structures associated with the
direct product of the tangent and cotangent bundle $T\oplus T^*$? This
question has recently been posed by Hitchin leading to the formulation of
generalized complex geometry \cite{hit}. Generalized complex geometry
naturally contains complex, symplectic, K\"ahler and bihermitian geometries
as particular cases. A fairly complete account can be found in Gualtieri's
thesis \cite{thesis}. Several geometric concepts, like the moment map, reduction
of generalized complex and generalized K\"ahler geometries, and others are
currently being developed \cite{bursztyn-crainic,crainic,lin-tolman,lin-tolman2,hu,bursztyn-cavalcanti
-gualtieri}.  A number of works have sought to clarify the connection of
generalized complex geometry to supersymmetry. A very interesting analysis
presented in \cite{lmtz} showed how the integrability conditions of the
generalized complex structure could be understood, at the nonlinear sigma
model level, as the conditions for a manifestly $(1,1)$
supersymmetric model to be $(2,2)$ supersymmetric. In the language of representations it has also
become increasingly evident that semi-chiral superfields play a central role
\cite{lmtz,lruz,lruz2,blp,blpz}.

{\it The role of semi-chiral superfields and our work:} The mathematical
literature seems to allow for some ambiguities in the definition of some of
the geometric structures involved. In particular, several groups have
proposed definitions of the reduction of  generalized complex geometry and
moment map \cite{lin-tolman,lin-tolman2,hu,bhc2}. On the physics side,
these concepts are related to gauging some of the symmetries of the
nonlinear sigma model. The above situation motivates us to approach these
concepts from the physics point of view. We also believe that understanding
nonlinear sigma models  and the gauging of some of the symmetries in the
most general context is an important problem in physics. We have heavily
relied on previous works that addressed such questions in the context of
complex and K\"ahler geometry and partially in the case of bihermitian
geometry  \cite{hps,hklr,bw,hs}. A concept that does not seem to be
intrinsic to the mathematical literature but that we will investigate,
following in part the work of  Ro\v{c}ek and Verlinde \cite{rv}, is T-duality
in the presence of semi-chiral superfields. More generally, the goal of this
paper is to  {\it lay the groundwork for exploring the connection between
generalized complex geometry and supersymmetry in terms of gauged nonlinear
sigma models with superfields in the semi-chiral representation.}

The paper is organized as follows.  In section 2 we discuss the gauging of
the sigma model with semi-chiral superfields under the simplifying assumption that the K\"{a}hler potential is
invariant under the action of the $U(1)$ symmetry.  We reduce the action to $(1,1)$
superspace and find, via comparison with \cite{hps}, the moment map and the
one form required for gauging a sigma model with $B$ field .  We analyze the
example of the $SU(2)\otimes U(1)$ WZNW model and verify the
identifications we have made for the moment map and one form and point out an
ambiguity which arises in  the presence of  semi-chiral
representations. We conclude section 2 with a description of a gauging based
on the prepotential. In section 3 we  briefly review the mathematical
literature on Hamiltonian action and moment map for generalized complex
geometry. We also compare the mathematical definition with the physical
definition of the moment map and explicitly discus the example of the
$SU(2)\otimes U(1)$ WZNW model.  In section 4 we address T-duality in (2,2)
superspace. We first describe T-duality in case of
chiral and twisted chiral superfields following the formalism used in
\cite{rv}, but with a different gauge fixing procedure.  Next, we use this
gauge fixing procedure to describe T duality in the case when semi-chiral
superfields are used.  We finish by working out the 4d flat space example.  In section 5 we draw some conclusions and point out
some interesting open questions.

\section{Gauging a (2,2) sigma model and the moment map}
\indent

One property of  $(2,2)$ superspace is that the  sigma model is given
entirely in terms of the K\"{a}hler potential:
\be
{\cal S}=\int d^2 x D^2 \bar D^2 K(\{\Phi,\bar\Phi\}).
\ee

Geometric quantities associated with the sigma model, such as the target space
metric and B field, can be obtained in explicit form by performing a reduction to $(1,1)$ superspace. The specific properties of the sigma model target
space geometry follow from the specific choice of $N=2$ superfields $\{\Phi\}$.
If $\{\Phi\}$ is a set of chiral superfields $\bar D_\pm \Phi=0$, or twisted chiral
superfields $D_+\Phi=\bar D_-\Phi=0$, then the associated target space
geometry is K\"ahler. If both chiral and twisted
chiral superfields are found among $\{\Phi\}$, then the corresponding geometry
goes by the name of ``bihermitian geometry'' with almost product structure.
This particular geometry is characterized by two {\it commuting}
almost complex structures $J_\pm$, $J_\pm^2=-1$, which are integrable and
covariantly constant with respect to affine connections with torsion, a metric
that is bihermitian with respect to both  complex structures
$J^t_\pm  g J_\pm =g$, and a $B$-field.
As a consequence of the fact that $[J_+,J_-]=0$, the
metric acquires a block diagonal form, inducing a natural decomposition of
the tangent space, along the chiral, respectively, twisted chiral components.
It is the feature of having two commuting  complex structures that
distinguishes the almost product structure geometry among the class of
bihermitian geometries.
Lastly, if the set of superfields which determine the K\"ahler potential contains
a more general $(2,2)$ semi-chiral superfield, constrained only by a
single superspace covariant derivative $D_+\Phi=0$, or $D_- \Phi=0$, then the
sigma model target space is bihermitian, with non-commuting 
complex structures. Recently, it has been shown \cite{thesis}
that the projection of the generalized K\"ahler geometry onto the tangent
bundle yields the bihermitian data $(g, J_\pm, B)$ found by Gates, Hull and
Ro\v{c}ek  \cite{ghr} when investigating non-linear sigma models with $(2,2)$
supersymmetry.

 If the sigma model target space has an isometry group, then a generic Killing
vector can be decomposed in a basis of the Killing vectors $k_A$ which generate
the Lie algebra of the isometry group
\be
\xi=\xi^A k_A=\xi^A k_A^i \partial_i, \qquad [k_A,k_B]= f_{AB}{}^C k_C,
\qquad {\cal L}_\xi g=0.
\ee
The infinitesimal transformation of the sigma-model fields is given by
\be
\delta \phi^i=\epsilon^A k_A^i,
\label{deltaphi}
\ee
where $\epsilon^A$ are rigid infinitesimal parameters.
For a sigma model with isometries, there are additional geometric data.
These follow from the integrability conditions associated with the
additional requirements that the action of the Killing vector leave
invariant not just the metric, but the field strength of the B-field $H$, and
the symplectic forms $\omega_\pm =g J_\pm$:
\be
\label{preserve}
{\cal L}_\xi H=0, \qquad {\cal L}_\xi \omega_\pm=0.
\ee
{From} the condition that $H$ is invariant, it follows that
\be
{\cal L}_\xi H=d i_\xi H+ i_\xi dH=d i_\xi H=0.
\ee
Since the two-form $i_\xi H$ is closed, locally it can be written as
\be
i_{\xi_A} H = du_A,
\ee
where the one-form $u$ is determined up to an exact, Lie-algebra valued
one-form. The ambiguity in $u$ can be fixed, up to $U(1)$ factors in the
Lie algebra, by requiring that it is equivariant
${\cal L}_A u_B=f_{AB}{}^C u_C$.

Besides this one form $u$, the other geometric datum associated with the
existence of an isometry group is the moment map
(also known as Killing potential).
From the condition that the symplectic form is invariant under $\xi$, and
from $d\omega_\pm (J_\pm X,J_\pm Y,J_\pm Z)=\pm H(X,Y,Z)$, it follows
that $\omega_\pm \xi \mp J_\pm^T u$ is closed. Therefore, locally one finds
\be
d\mu_\pm = \omega_\pm \xi \mp J_\pm^T u\label{mommap},
\ee
where $\mu_\pm$ are the moment maps. This expression is the generalization for
a manifold with torsion of the integrability condition for a Hamiltonian 
vector field $\xi$. Since $\xi$ is also Killing, it follows that ${\cal L}_\xi J_\pm=0$.

The relevance of these two quantities, the one-form $u$ and the moment map
$\mu$, becomes obvious when constructing the gauged sigma model,
by promoting the rigid (global) isometries $\ref{deltaphi}$ to local ones.
This is accompanied, in the usual manner, by introducing a compensating
connection (gauge potential) $\partial_\mu \phi^i
\longrightarrow \nabla_\mu\phi=\partial_\mu \phi^i + A_\mu^A k_A^i$,
which transforms as
$\delta A_\mu^A=\partial_\mu \epsilon^A+f_{AB}{}^C A_\mu^B\epsilon^C$.
For $(1,0)$ or $(1,1)$ supersymmetric sigma models, the bosonic gauge
connection becomes part of a corresponding $(1,0)$ or $(1,1)$ super Yang-Mills
multiplet.
Promoting the partial derivatives to gauge covariant derivatives is not
enough in the presence of a $B$-field \cite{hps ,hs,hs1}.
New terms, which depend on
the $u$ one-form and the moment map, must be added to
the sigma model action. For a bosonic, $(1,0)$ or $(1,1)$
supersymmetric sigma-model, adding only $u$-dependent terms suffices:
\be
{\cal S}=\int d^2 x d^2\theta
\bigg( g_{ij}\nabla_+\phi^i \nabla_-\phi^j
+B_{ij}D_+\phi^j D_-\phi^j
-2 u_{iA} A_{(+}^A D_{-)}\phi^i+ A_+^A A_-^B c_{[AB]}\bigg),
\label{11action}
\ee
where
$D_\pm$ are flat superspace covariant derivatives and $\nabla_\pm$
are superspace gauge covariant derivatives, while $c_{[AB]}=k_{[A}^i u_{iB]}$.

When the sigma model has additional supersymmetries,
then the gauged sigma model
action acquires new terms, which are moment map dependent.
The additional  supersymmetries, which at the level of
$N=1$ superspace are nonlinearly realized, are of the form
\be
\delta \phi^i=\epsilon J^i{}_j D_+\phi^j.
\ee
The supersymmetry algebra requires that the $(1,1)$ tensor(s) $J^i{}_j$
be identified with the almost complex structure(s).
The gauged (2,2) sigma model action typically contains a term
\be
\delta {\cal{S}}=\int d^2 x d^2 \theta S\mu,
\ee
where $\mu$ is the moment map, and $S$ is a super-curvature that
appears in the (2,2) superalgebra (more precisely in the super-commutators
$\{\nabla_+,\bar \nabla_-\}$).

The above formulae show that the moment map and the one-form $u$
are necessary ingredients to gauge the sigma model.
Intuitively, the one-form $u$ is
needed  to gauge the sigma model with $B$ field and the moment map is
needed to gauge a sigma model with extended supersymmetry.

Alternatively, one could choose instead to remain at the level of
$(2,2)$ superspace and perform the gauging there, without ever descending
onto the $(1,1)$ superspace. In the process,
the K\"ahler potential of the ungauged  sigma model $K(\{\Phi\})$ acquires
a new term, which is also moment-map dependent \cite{hps,hklr}.

We shall be interested in gauging $(2,2)$ sigma models whose target space
has a bihermitian structure, with non-commuting  complex structures.
The natural starting point for us is the $N=2$ superspace formulation of
 a sigma-model written in terms of $(2,2)$ semi-chiral superfields:
\bea
\label{semichiralsuperfieldsconditions}
&&\mbox{left chiral:}\qquad \bar D_+X=0,\cr
&&\mbox{right antichiral:}\qquad D_-Y=0.
\eea
We begin by making the observation that the following transformations are
consistent with the contraints\footnote{This is not the most general set of
transformations consistent with the constraints on $X$ and $Y$. However,
these are the only transformations relevant to our considerations.}
on $X$ and $Y$.
\bea
&&X\ra(A+B)X+C+D,\cr
&&Y\ra(F+G)Y+W+Z,
\label{gaugetransform}
\eea
where $A,C$ are chiral  superfields, $B,D$ are twisted anti chiral superfields, $F,W$ are anti chiral, and $G,Z$ are twisted chiral.
When these transformations correspond to gauge transformations they can
be properly accounted for using both the chiral and twisted chiral vector
multiplets.

For simplicity we will consider only the gauge transformations where the
semi-chiral superfields are multiplied and shifted by
chiral and anti-chiral superfields\footnote{We thank S. Gates for various clarifications on this point.}.
In this paper we shall follow two complementary  approaches
to constructing the gauged action in $(2,2)$ superspace.  The first
method involves descending at the level of $(1,1)$ superspace by
following the usual route of substituting the Grassmann integration by
differentiation $\int d \theta d\bar \theta\to D\bar D$, and
by the subsequent replacement of  the ordinary superspace covariant derivatives
by gauge covariant derivatives $D\bar D \to \nabla\bar\nabla$. This is
equivalent to gauging by minimal coupling, if the K\"ahler potential
is invariant under the action of the isometry generators.
The second method \cite{hklr} uses
the prepotential of the gauge multiplet $V$ explicitly in the K\"ahler potential
to restore the invariance of the action under local transformations.

For simplicity we restrict ourselves to  $U(1)$ isometries.
As such we can go to a coordinate system where the isometry is realized
by a shift of some coordinate. This  implies that the K\"{a}hler
potential $K(X,\bar X, Y,\bar Y)$ will be independent of a certain
linear combination of the left and right semi-chiral superfields.
For example, for
\be
\label{isok}
K=K(X+\bar X, Y+\bar Y, X+Y).
\ee
we can immediately read off the Killing vector associated with the
isometry.  In this case it takes the form
\be
\xi=i\frac{\partial}{\partial X}-i\frac{\partial}{\partial \bar X}-i\frac{\partial}{\partial Y}+i\frac{\partial}{\partial \bar Y}.
\ee
{From} (\ref{gaugetransform}) we see that this is an example of a K\"ahler
potential, with a U(1) isometry which can
be gauged using only the (un-twisted) (2,2) super Yang-Mills multiplet.

\subsection{Gauging and the reduction to (1,1) superspace}
\indent

Let us consider the first of the two approaches to gauging which we have
outlined before. Since we are interested in extracting the geometric data
(including those associated with isometries) from the sigma model, and
these are most easily seen in the language of $(1,1)$ superspace, here we
describe the bridge from $(2,2)$ to $(1,1)$ superspace, following \cite{ghr} closely.

We begin by recording the $(2,2)$ gauge covariant supersymmetry algebra for
the (un-twisted) super Yang-Mills multiplet $(A_\alpha^A, \bar A^A_\a, A_\pp^A,
A_{\mm}^A)$:
\bea
\label{the22algebracovariant}
[\nabla_\a,\nabla_\b\}&=&0, \cr
[\nabla_\a,\bar\nabla_\b\}&=&2i(\g^c)_{\a\b}\nabla_c+
2g[C_{\a\b}S-i(\g^3)_{\a\b}P]\x,\cr
[\nabla_\a,\nabla_b\}&=&\lambda(\g_b)_\a{}^\b \bar W_\b \x,\cr
[\nabla_a,\nabla_b\}&=&-i\lambda\epsilon_{ab}\mathcal{W}\x,
\eea
where the bosonic two-dimensional indices are $a,b,c=\{\pp,\mm\}$,  and
the Grassmann odd two-dimensional spinor indices are $\a=\pm$. The skew-symmetric tensor $C_{\a\b}$ is  used for
raising and lowering indices in superspace. Having in mind the gauging of a
certain isometry of a (2,2) sigma model, we used the Killing vector $\x$
to denote the couplings of the sigma model superfields to the
(2,2) super Yang-Mills multiplet.

Also, following from the Bianchi identities, one has
the following set of constraints 
\bea
\nabla_\a S&=&-i\bar W_\a, \qquad \nabla_\a P=-(\g^3)_\a{}^\b\bar W_\b,\cr
\nabla_\a \bar W_\b&=&0, \qquad \nabla_\a d= (\g^c)_\a^\b\nabla_c \bar W_\b, \cr
\nabla_\a W_\b &=& iC_{\a\b}d-(\g^3)_{\a\b}\mathcal{W}+(\g^a)_{\a\b}\nabla_a S-i(\g^3\g^a)_{\a\b}\nabla_a P.
\eea

According to our
previous discussion on gauging methods, we take our first step towards
constructing the gauged (2,2) sigma model begin by making the substitution
\be
\label{measure}
\int d^2\bar\q d^2\q=\frac{1}{8}[\nabla^\a\nabla_\a\bar\nabla^\b\bar\nabla_\b+
\bar\nabla^\b\bar\nabla_\b\nabla^\a\nabla_\a],
\ee
where we have used  the conventions of \cite{gates}.

In order to reduce the action written in $(2,2)$ superspace to
$(1,1)$ superspace we need to express the $(2,2)$ gauge covariant
derivatives in terms of two copies of the $(1,1)$ derivatives:
\be
\hat \nabla_\a=\frac{1}{\sqrt{2}}(\nabla_\a+\bar \nabla_\a), \qquad \tilde
\nabla_\a=\frac{i}{\sqrt{2}}(\nabla_\a-\bar \nabla_\a).
\ee
The $(1,1)$  derivatives satisfy the following algebra:
\bea
[\hat \nabla_\a,\hat \nabla_\b\}&=&2i(\g^c)_{\a\b}\nabla_c-2i\lambda(\g^3)_{\a\b}P
\xi,\cr
[\hat \nabla_\a,\nabla_b\}&=&\frac{\lambda}{\sqrt{2}}(\g_b)_\a{}^\b\hat W_\b \xi,\cr
[\tilde \nabla_\a,\tilde \nabla_\b\}&=&2i(\g^c)_{\a\b}\nabla_c-
2i\lambda(\g^3)_{\a\b}P\xi,\cr
[\tilde \nabla_\a,\nabla_b\}&=&\frac{\lambda}{\sqrt{2}}(\g_b)_\a{}^\b \tilde W_\b
\xi,\cr
[\hat \nabla_\a,\tilde \nabla_\b\}&=&-2i\lambda C_{\a\b}S\xi,
\eea
where $\tilde W_\b=\frac{i}{\sqrt{2}}(\bar W_\b-W_\b)$ and
$\hat W_\b=\frac{1}{\sqrt{2}}(\bar W_\b+W_\b)$.

Next we consider the measure of the $(2,2)$ action (\ref{measure}) and
we rewrite it in terms of
the $(1,1)$ derivatives
\be
\hat\nabla^\a\hat\nabla_\a\tilde\nabla^\b\tilde\nabla_\b=2\nabla^\a\nabla_\a
\bar\nabla^\b\bar\nabla_\b+2\bar\nabla^\b\bar\nabla_\b\nabla^\a\nabla_\a
+(...)\xi+\mbox{total derivative}
\ee
Therefore $\hat\nabla^\a\hat\nabla_\a\tilde\nabla^\b\tilde\nabla_\b$
and $2\nabla^\a\nabla_\a\bar\nabla^\b\bar\nabla_\b+2\bar\nabla^\b\bar\nabla_\b
\nabla^\a\nabla_\a$ are equivalent when acting on a potential which is
invariant under the isometry, that is,  satisfies $\xi K=0$.

Reducing the $(2,2)$ Lagrangian amounts to evaluating
\be
L=\int d^2\bar\q d^2\q K=\frac{1}{4}\hat\nabla^2\tilde\nabla^2 K(X,\bar X, Y, \bar Y),
\ee
where the semi-chiral superfields $X,\bar{X},Y, \bar{Y}$ obey (\ref{semichiralsuperfieldsconditions}), but with 
the ordinary superspace derivative $D_{\pm}$ replaces by the gauge-covariant derivatives (\ref{the22algebracovariant}).
Then, using the relation
\be
\tilde\nabla^\a\tilde\nabla_\a=-2i\tilde\nabla_+\tilde\nabla_- -2i\lambda P\xi,
\ee
we only need to evaluate $\tilde\nabla_+\tilde\nabla_- K$.
Additionally, we must decompose the (2,2) left and right semi-chiral
superfields into (1,1) superfields
\be
\varphi= X|,\qquad \Psi= \tilde \nabla_- X|, \qquad
\chi = Y|,\qquad \Upsilon= \tilde\nabla_+ Y|.
\ee
We end up with:
\bea
\label{reduction}
\tilde\nabla_+\tilde\nabla_-K&=&\hat\nabla_+
\varphi^Im_{II'}\hat\nabla_-\chi^{I'}
+\Upsilon_+^{I'}n_{I'I}\Psi_-^I
+\Psi_-^I(2\o_{IJ}\hat\nabla_+\varphi^J+ip_{II'}\hat\nabla_+\chi^{I'})\cr
&~&+\Upsilon_+^{I'}(2\o_{I'J'}\hat\nabla_-\chi^{J'}-iq_{I'I}\hat\nabla_
-\varphi^I)
-2i\lambda SK_{i'}\xi^{i'}+2igSK_{\bar i'}\x^{\bar i'}\cr
&~&+2i\lambda (S+iP)K_i\x^i-2i\lambda (S-iP)K_{\bar i}\x^{\bar i}\cr
&=&\hat\nabla_+\Phi^T \cdot E\cdot \hat\nabla_-\Phi+S_{+I}u^{II'}S_{-I'}-
2i\lambda SK_{i'}\x^{i'}+2i\lambda SK_{\bar i'}\x^{\bar i'}\cr
&~&+2i\lambda (S+iP)K_i\x^i-2i\lambda (S-iP)K_{\bar i}\x^{\bar i},
\eea
where we have used the notation
$K_i=\partial_{\varphi^i}K, K_{i'}=\partial_{
\chi^{i'}}K$ and that the U(1) Killing vector is $\xi=\xi^i\partial_{\varphi^i}
+\xi^{i'}\partial_{\chi^{i'}}+\xi^{\bar i}\partial_{\bar\varphi^{\bar i}}
+\xi^{\bar i'}\partial_{\bar\chi^{\bar i'}}$. The index $I$ is a collective
index: $I=\{i,\bar i\}$, and $\Phi=\{\phi,\bar\phi,\chi,\bar\chi\}$.
The matrices $m,n,\o,p,q$, expressed in terms of the second order
derivatives of the K\"ahler potential, are the same as in \cite{lruz}. Also,
analogous to \cite{lruz}
\bea
S_{+I}u^{II'}&=&\Upsilon_+^{I'}-2u^{II'}\o_{IJ}\hat\nabla_+\varphi^J-
iu^{II'}P_{IJ'}\hat\nabla_+\chi^{J'}\cr
u^{II'}S_{-I'}&=&\Psi_-^A+2u^{II'}\o_{I'J'}\hat\nabla_-\chi^{J'}-
iu^{II'}q_{I'J}\hat\nabla_-\varphi^B\cr
E&=&g+B\;=\;\left(
\begin{array}{cc}
2i\o uq&m-4\o u\o'\\
p^tuq&2ip^tu\o'\\
\end{array}
\right).\label{E}
\eea
At a first glance it appears that we have an asymmetric coupling of the field
strength $P$ between the fields $\varphi^I$ and $\chi^{I'}$.
However, this is just an artifact of our choice in evaluating the
covariant derivatives.  Note that
\bea
\xi K(\varphi,\chi)=0\ra K_i \xi^i+K_{\bar i}\x^{\bar i}
+K_{i'}\x^{i'}+K_{\bar i'}\x^{\bar i'}=0.
\eea

This means that the reduced Lagrangian is given by
\bea
\label{redox}
L&=&\hat \nabla^\alpha\hat\nabla_\alpha\bigg(
\hat\nabla_+\Phi^T \cdot E \cdot \hat\nabla_-\Phi+S_{+I}u^{II'}S_{-I'}\cr
&~&+2i\lambda (S+\frac{i}{2}P)K_i\x^i-
2i\lambda (S-\frac{i}{2}P)K_{\bar i}\x^{\bar i}\cr
&~&-2ig(S+\frac{i}{2}P)K_{i'}\x^{i'}+
2ig(S-\frac{i}{2}P)K_{\bar i'}\x^{\bar i'}\bigg)\cr
&=&\hat \nabla^\alpha\hat\nabla_\alpha\bigg(
\hat\nabla_+\Phi^T \cdot (g+B)\cdot \hat\nabla_-\Phi+S_{+I}u^{II'}S_{-I'}\cr
&~&+2i\lambda S(K_i\x^i-K_{\bar i}\x^{\bar i}-
K_{i'}\x^{i'}+K_{\bar i'}\x^{\bar i'})\cr
&~&-\lambda P(K_i\x^i+K_{\bar i}\x^{\bar i}-
K_{i'}\x^{i'}-K_{\bar i'}\x^{\bar i'})\bigg).
\label{mainresult}
\eea
This is the gauged sigma model we were after, and  it is
one of our main results.

To understand the various terms that appear in (\ref{mainresult}), it is
useful to compare this action with (\ref{11action}), given that both
actions represent gauged sigma models with manifest (1,1)
supersymmetry.
This explains the obvious common elements $\hat\nabla^\alpha\Phi^T \cdot g
\cdot\hat \nabla_\alpha\Phi + \hat D^\alpha \Phi^T \cdot B\cdot
\hat D_\alpha\Phi$. The gauging of the $B$-field terms is done in
(\ref{11action}) by including the $u$-dependent terms. How about in our case?
First, we notice that since we have assumed that $\xi K=0$, in other words that the
minimal coupling prescription will suffice, this is indeed what the
gauged sigma model Lagrangian (\ref{mainresult}) reflects.
The extra terms required for the gauging of the $B$-field terms
can be combined into $i_\xi B\cdot \hat D_{(-}\Phi A_{+)}$.
As a consequence of the assumption $\xi K=0$, we find that
$\mcal L_\xi B=0$. This is a stronger condition than $\mcal L_\xi H=0$,
and it implies the latter. Since $\mcal L_\xi B=0$, we find
\be
u=-i_\xi B+ d\sigma,
\ee
where $d\sigma$ is an exact one-form invariant under the action of the
isometry group.
This is exactly what is required to match the minimal coupling of the
$B$-field terms against the $u$-terms
in (\ref{11action}).  The $c_{AB}$ terms in (\ref{11action}) vanish
in the case of a $U(1)$ gauging. Otherwise, they, too, could be recognized
in the minimal coupling gauging of (\ref{11action}).

We shall see that the ambiguity in defining $u$, namely
the exact one-form $d\sigma$, is reflected in  (\ref{mainresult}) in the term
which multiplies the field strength $P$. The expression
$-\lambda d(K_I \xi^I-K_{I'}\xi^{I'})$ is $d(\sigma)$. We verify that it is invariant
under the U(1) action:
\bea
\mcal L_\xi d\sigma&=&d(i_\xi d\sigma)=d\bigg(
(\xi^I\partial_I+\xi^{I'}\partial_{I'})(
\xi^J\partial_J-\xi^{J'}\partial_{J'})K\bigg)=\nonumber\\
&=&2d\bigg(
(\xi^I\partial_I+\xi^{I'}\partial_{I'})\xi^I\partial_J \;K\bigg)=
2d\bigg((
-\xi^I\partial_I \xi^{I'}\partial_{I'}+\xi^{I'}\partial_{I'}\xi^I\partial_I)
K\bigg)=0,\nonumber\\
\eea
 where in the last step we used that we can go to a coordinate system where
the U(1) action is realized by a shift of some coordinate, which
implies $[\xi^I\partial_I ,\xi^{I'}\partial_{I'}]=0$.

The remaining terms in (\ref{mainresult}), such as those dependent on the
 auxiliary superfields $S_\pm$ and which have no counterpart in
(\ref{11action}), are present because our starting point was
a (2,2) supersymmetric action with off-shell (2,2) superfields.
Lastly, we recognize in the terms proportional to the superfield strength
$S$, a linear combination of the moment maps.
Their presence is required
to insure the invariance of the gauged sigma model action.
While the expression proportional to S in (\ref{redox}) is not immediately
relatable to the moment map given in (\ref{mommap}), it does have a form
similar to that given in \cite{agf,hps,Jourjine:1983xs} for the
moment map.  There the moment map is identified as the imaginary part of the
holomorphic transformation of the K\"{a}hler potential under the action of the
Killing vector.

Thus we conclude with the identifications:
\bea
&&\mbox{Moment map}\sim ~K_i\x^i-K_{\bar i}\x^{\bar i}-
K_{i'}\x^{i'}+K_{\bar i'}\x^{\bar i'}\label{holok}\\
&&\s\sim ~K_a\x^i+K_{\bar i}\x^{\bar i}-
K_{i'}\x^{i'}-K_{\bar i'}\x^{\bar i'}.
\eea
These identifications, and especially  the rapport between  (\ref{holok})
and (\ref{mommap}), will be verified in the next section.

\subsection{An example: the $SU(2)\times U(1)$ WZNW model}
\indent

In this section we apply our previous construction of a (2,2) gauged sigma
model to a concrete example: the  $SU(2)\times U(1)$ WZNW model.
The $(2,2)$ supersymmetric $SU(2)\times U(1)$ WZNW sigma model was first
formulated in terms of semi-chiral superfields in \cite{ikr}.  These 
authors discovered non-commuting complex structures on $SU(2)\times U(1)$
and constructed a duality functional that does not change the geometry.
However, this duality functional allows to map between two seemingly
different descriptions, one for $SU(2)\times U(1)$ described in terms of
chiral and twisted chiral superfields and the another description in
terms of semi-chiral superfields.  The explicit form of the K\"ahler potential
was given in \cite{st, gmst}. A discussion on the various dual descriptions
which can be obtained by means of a Legendre transform can be found in
\cite{gmst}. The $SU(2)\times U(1)$ K\"ahler potential is
\be
K=-(\bar\phi+\eta)(\phi+\bar\eta)+\frac{1}{2}(\bar\eta+\eta)^2-
2\int^{\bar\eta+\eta}dx\mbox{ln}(1+e^{x/2}),
\ee
where $\bar D_+\phi=D_-\eta=0$.  Because $K=K(\bar\phi+\eta, \phi+\bar\eta,
\eta+\bar \eta)$ we cannot directly gauge the theory, using only
the coupling with the (un-twisted) (2,2) super Yang-Mills multiplet.
However, there is an easy remedy to this problem, namely we shall use a dual
description, found via a Legendre transform \cite{gmst}:
\be
K(r,\bar r,\eta,\bar \eta)=K(\phi, \bar\phi,\eta,\bar \eta)-r\phi-\bar r\phi,
\ee
where $r$ is semi-chiral, $\bar D_+ r=0$, and $\phi$ is unconstrained.
By integrating over $r$, we recover the previous K\"ahler potential.
On the other hand, by integrating over $\phi$, that is eliminating it from
its equation of motion, we find a K\"ahler potential
$K=K(r+\eta,\bar r+\bar\eta, \eta+\bar\eta)$ (up to terms that represent a
generalized K\"ahler transform $\frac 12\eta^2+\frac 12\bar \eta^2$). This
is an example of a ``duality without isometry'' \cite{gmst},
where the K\"ahler potential of a  semi-chiral superfield sigma model can be
mapped via Legendre transforms into four different, but equivalent expressions,
all involving only semi-chiral superfields.

The new form taken by the $SU(2)\times U(1)$ K\"ahler potential
\be
\tilde K=(\bar r+\bar \eta)(r+\eta)-2\int^{\bar\eta+\eta}dx\mbox{ln}(1+e^{x/2})
\ee
indicates that the  $U(1)$ isometry is realized by the transformations
\be
r\ra r+i\e,~~\eta\ra\eta + \overline{(i\e)},
\ee
where $\epsilon$ is a constant real parameter.
However, when promoting this symmetry to a local one, according to our previous discussion, $\epsilon$ is to be
interpreted as a chiral superfield, and $\bar\epsilon$ as an anti-chiral
superfield.

The K\"ahler potential is left invariant under the action of the (2,2)
Killing vector
\be
\xi=i\frac{\partial}{\partial r}-i\frac{\partial}{\partial \bar r}-i\frac{\partial}{\partial \eta}+i\frac{\partial}{\partial \bar\eta}.
\ee
{From} (\ref{E}) we can now calculate the B field, its field strength and
their contractions with the Killing vector:
\bea
\label{stuff}
&&B=(1-2f)(dr\wedge d\bar\eta+d\bar r\wedge d\eta)\cr
\cr
&&\mbox{i}_\xi B=i(1-2f)d\bar\eta-i(1-2f)d\eta+i(1-2f)d\bar r-i(1-2f)dr\cr
\cr
&&H=dB=2(\frac{\partial f}{\partial\eta}dr-\frac{\partial f}{\partial\bar\eta}d\bar r)\wedge d\eta\wedge d\bar\eta\cr
\cr
&&\mbox{i}_\xi H=d(2if[-dr+d\bar r-d\eta+d\bar\eta])=du,
\eea
where
\bea
&&f=f(\bar\eta+\eta)=\frac{\mbox{exp}[\frac{1}{2}(\bar\eta+\eta)]}
{1+\mbox{exp}[\frac{1}{2}(\bar\eta+\eta)]}.
\eea
We also find that $\mcal{L}_\xi B=0$, in accord to the expectation that
the gauging is done via minimal coupling \cite{hs,hs1}.
As discussed before, it implies that $u=-\mbox{i}_\xi B+d\s$
where $d\s$ is an exact one-form, invariant under the action of the Killing
vector.
As to the term proportional to $P$ in (\ref{redox}) we find that is equal to
$2i\lambda \s$, where $d\sigma=d(\bar r-r+\bar \eta-\eta)$. Indeed, this one-form
satisfies the condition $i_\xi d\s=0$.

Next, we show how the term proportional to $S$ corresponds to the moment map.
\subsubsection{The Moment Map}
\indent

Here we verify that the term proportional to the super-curvature $S$
in (\ref{mainresult})
\be
i(K_{r}-K_{\bar r}-K_\eta+K_{\bar\eta})=2i
\bigg
[r+\bar r +\eta+\bar\eta-2\mbox{ln}(1+\mbox{exp}(\frac{\eta+\bar\eta}{2}))
\bigg]\equiv M\label{M}
\ee
is a certain linear combination of the two
moment maps of the bihermitian geometry. We recall their definition
\be
\label{sympmap}
g_{ij}\xi^j\pm u_i=I_{\pm}^j{}_i\del_j\mu_{\pm}.
\ee
Before we consider (\ref{sympmap}) we must first address the
ambiguity in the expression for the one form $u$.
The one-form $u$ is defined only up to an exact one form that
satisfies $\mcal{L}_\xi d\s=0$:
$
u=2if[-dr+d\bar r-d\eta+d\bar\eta]+ di(C_r r+C_{\bar r} \bar r+
C_\eta \eta+C_{\bar\eta}\bar\eta)
$
with $C_{r,\bar r,\eta,\bar\eta}$ constants, constrained only by
$C_r-C_{\bar r}-C_\eta+C_{\bar \eta}=0$.
However, our previous considerations have eliminated most of the
freedom in $d\sigma$, given that, from the gauged action we have identified
$C_r=-1, C_{\bar r}=1, C_{\eta}=-1,C_{\bar\eta}=1$.
Armed with the concrete expressions of the moment maps
we find the following relationship with $M$:
\be
M=-( \mu_+ + \mu_-).
\ee

\subsection{Alternative gauging procedure: the prepotential}
\indent

In section 2.1 we gauged the sigma model by replacing the Grassmann integration
measure with gauge supercovariant
derivatives and thus reducing the (2,2) action to a gauged action with (1,1)
manifest supersymmetry.
Here we take the alternative approach of using the gauge prepotential
superfield $V$ to arrive at a gauge-invariant K\"ahler potential.
This procedure is done in (2,2) superspace, and all supersymmetries
remain manifest. Therefore this gauging method has the advantage of
facilitating the discussion of duality functionals, which we will address
in the next section.

In simple cases, the gauging is done by adding the prepotential $V$ to the
appropriate combination of superfields in the K\"ahler potential: for the
example $K=K(X+\bar X, Y+\bar Y, X+Y)$, the global symmetry is promoted to
a local one via the substitution 
\be
K(X+\bar X, Y+\bar Y, X+Y)\ra K(X+\bar X+V, Y+\bar Y+V, X+Y+V),
\ee
if the gauging is done using the (un-twisted) (2,2) super Yang-Mills
prepotential, in other words, if the gauge parameter is a chiral
superfield. On the other hand, if the gauge parameter is a
twisted chiral superfield, then we must use  the gauge prepotential
associated with a twisted (2,2) super Yang-Mills multiplet $V_t$. For
example, we could gauge $K=K(X+\bar X, Y+\bar Y, X+\bar Y)$ by
\be
K(X+\bar X, Y+\bar Y, X+\bar Y)\ra K(X+\bar X+V_t, Y+\bar Y+V_t, X+\bar Y+V_t).
\ee

For concreteness we continue to address only the gauging done using the
coupling to the (un-twisted) (2,2) super Yang-Mills multiplet.
In general, the isometry transformations of a given superfield are given by:
\be
X\ra e^{i\e \xi}X\Ra\bar X\ra e^{-i\bar \e \xi}\bar X,
\ee
where $\xi$ denotes the isometry generator and $\epsilon$ is a real
valued constant parameter.
When promoting this global symmetry to a local one, the gauge parameter
$\epsilon$ becomes a chiral superfield, and $\bar \epsilon$  an anti-chiral
superfield. The invariance is restored
by introducing the gauge prepotential superfield $V$,
 transforming as
 \be
V\ra V+i(\bar\e-\e).
\ee
We include $V$ through the replacement:
\be
\bar X \ra e^{iV\xi}\bar X.\label{gauge1}
\ee
Now $\bar X$  transforms in the same way as in the global case
and thus the invariance has been restored.

Although we have used the whole Killing vector $\xi$ in constructing the field
that transforms properly (\ref{gauge1}), to be more specific,
it is only the part of the Killing vector that
induces a transformation with the anti-chiral gauge parameter
which contributes to this definition.
In the example that we gave, $K=K(X+\bar X, Y+\bar Y, X+Y)$, $X,\bar Y$
transform with a chiral gauge parameter, and $\bar X, Y$, with an anti-chiral
parameter. The Killing vector will generally
factorize  $\x=\x_c+\x_{\bar c}$ such that $\x_c$ and $\bar \x_c$
induce a chiral
parameter, respectively an anti-chiral parameter gauge transformation.
In the $SU(2)\otimes U(1)$ example we have
$\x_{\bar c}=-i\frac{\partial}{\partial \bar r}
-i\frac{\partial}{\partial \eta}$.

Therefore we define
\be
\tilde X=e^L\bar X, ~~L=iV\x_{\bar c}.
\ee
The new field, $\tilde X$,
transforms under the gauge transformation in the exact same way
as $\bar X$ did under the global isometry.
Therefore by replacing $\bar X$ in the K\"ahler
potential  by $\tilde X$ we insure that the transformation of the K\"ahler
potential under the local transformation is the same as for the global
isometry, namely it is a generalized K\"ahler transformation. Of course, the
other semi-chiral superfield $Y$ undergoes a similar treatment:
\be
\tilde Y=e^L Y.
\ee

If the  K\"{a}hler potential  remains invariant under the action of the
Killing vector i.e. $\xi K(X,\bar X,Y,\bar Y)=0$, the
minimal coupling perscription is given by replacing $\bar X$ with
$\tilde X$ and $Y$ with $\tilde Y$.  Specifically, the gauged (2,2) Lagrangian
is given by the replacement
\be
K(X,\bar X,Y,\bar Y)\ra K(X,\tilde X, \tilde Y,\bar Y).
\ee
At this point we can use the relation
$K(X,\tilde X, \tilde Y,\bar Y)=e^LK(X,\bar X,Y,\bar Y)$
to rewrite the Lagrangian as
\bea
\label{geda}
K(X,\tilde X, \tilde Y,\bar Y)&=&e^LK(X,\bar X,Y,\bar Y)=K(X,\bar X,Y,\bar Y)
+\frac{e^L-1}{L}LK\cr
&=&K(X,\bar X,Y,\bar Y)+\frac{e^L-1}{L}VM,
\eea
where in $M=i\x_{\bar c}K$ we recognized the same object which we have
identified from the gauged (1,1) action as the moment map
(\ref{M}).

Next, we address the case of a K\"{a}hler potential which under the
action of the isomtery generator transforms with terms that take
the form of generalized K\"ahler transformations
\be
\x K=f(X)+\bar f(\bar X)+g(Y)+\bar g(\bar Y).
\ee
The trick is to introduce new coordinates and add them to the K\"{a}hler
potential in such a way that the new K\"{a}hler potential is invariant
under the transformation generated by the new Killing vector.
Specifically we introduce $\a,\b$ with $\bar D_+\a=D_-\b=0$.
We construct the new K\"{a}hler potential and Killing vector
\bea
\label{newstuff}
&&K'(X,\bar X, Y,\bar Y,\a,\bar\a,\b,\bar\b)=
K(X,\bar X,Y,\bar Y)-\a-\bar\a-\b-\bar\b \cr
\cr
&&\x'=\x+f(X)\frac{\del}{\del\a}+
\bar f(\bar X)\frac{\del}{\del\bar\a}+g(Y)\frac{\del}{\del\b}+
\bar g(\bar Y)\frac{\del}{\del\bar\b}.
\eea
Now the new K\"{a}hler potential $K'$ is invariant under the new Killing
vector
$\mcal L_{\x'} K'=0$ and we can proceed as before.
We replace all fields which transform with the parameter $\bar\e$
with the combination which transforms with the field $\e$ by using
$e^{L'}$ where  $L'=iV\x'_{\bar c}$.
Next we define the tilde versions of $\bar X, Y, \bar\a, \b$ as follows
\bea
\tilde X=e^{L'}\bar X,\qquad
\tilde Y=e^{L'} Y,\qquad
\tilde \a=e^{L'}\bar\a\qquad
\tilde \b=e^{L'}\b.
\eea
The gauged Lagrangian is obtained by the same substitution as before.
Finally we get
\bea
K'(X,\tilde X,\tilde Y,\bar Y,\a,\tilde\a,\tilde\b,\bar\b)&=&K(X,\tilde X,\tilde Y,\bar Y)-\a-\tilde\a-\tilde\b-\bar\b\cr
&=&e^LK(X,\bar X,Y,\bar Y)-i\frac{e^L-1}{L}V(\bar f(\bar X)+g(Y))\cr
\cr
&=&K(X,\bar X,Y,\bar Y)+\frac{e^L-1}{L}(LK-iV\bar f(\bar X)-iVg(Y))\cr
\cr
&=&K(X,\bar X,Y,\bar Y)+\frac{e^L-1}{L}VM.
\eea

\section{Eigenspaces of Generalized Complex Structures}

\subsection{Hamiltonian action and moment map in the mathematical literature}
\indent

In the context of generalized complex geoemtry,
the origin of subsequent definitions of the Hamiltonian action can be found
in  Gualtieri's thesis \cite{thesis} where it
was shown that certain infinitesimal symmetries preserving the
generalized complex structure ${\cal J}$ can be
extended to second order.

Intuitively, given a Hamiltonian action on a generalized complex manifold, the moment map is a quantity
that is constant along the action of the group elements.
More formal definitions of moment
map were given, for example, in \cite{lin-tolman, lin-tolman2,hu, hu2};
in \cite{hu}  Hu considered the Hamiltonian group globally.
For concreteness here we will explore one of the definitions put
forward by Lin and Tolman \cite{lin-tolman} in the simplest
setting without $H$-twisting, namely, definition 3.4:

{\it
Let a compact Lie group $G$ with Lie algebra $g$ act on a manifold $M$, preserving a generalized complex
structure ${\cal J}$. Let $L\in T\oplus T^*$ denote the
$\sqrt{-1}$-eigenbundle of ${\cal J}$. A generalized moment map
is a smooth function $\mu: M\to  g^*$ so that

(i) $\xi_M-\sqrt{-1} \,d\mu^{\xi}$ lies in  $L$ for all $\xi\in g$,
where $\xi_M$ denotes the induced vector field on M.

(ii) $\mu$ is equivariant.
}

In subsequent works, the definition of Hamiltonian
action was generalized to include the $H$-twisted case \cite{lin-tolman2,hu}.
In \cite{bursztyn-cavalcanti-gualtieri}, the authors  arrived at a
definition of moment map in terms of the action of a Lie algebra on a Courant algebroid.

In what follows we will explore the particular definition cited above, and
compare it with the expressions that we gave for the moment map in the
previous sections.
We leave for a future publication the issue
of the equivalence of the various definitions given in the math literature,
and their relationship with the physical point of view advocated in this paper,
via the gauging of the (2,2) sigma model.

\subsection{Generalized Kahler geometry and the eigenvalue problem}
\indent

In a series of papers \cite{lruz,lruz2} the authors established that chiral, twisted chiral, and
semi-chiral superfields are the most generic off-shell multiplets for  $\mcal{N}=(2,2)$ supersymmetric
non-linear sigma models. The use of these $(2,2)$ multiplets yields generalized K\"{a}hler geometries.

To practically use the above definition of moment map in the case of K\"ahler geometry we recall that,
according to  Gualtieri (see Chapter 6 in  \cite{thesis}), the generalized
complex structures of the generalized K\"ahler geometry take the following
expressions:
\be
\mcal{J}_{1/2}=
\frac{1}{2}\left(
\begin{array}{cc}
1&0\\
B&1\\
\end{array}
\right)
\left(
\begin{array}{cc}
J_+\pm J_-&-(\o_+^{-1}\mp \o_-^{-1})\\
\o_+\mp \o_-&-(J_+^t\pm J_-^t)\\
\end{array}
\right)
\left(
\begin{array}{cc}
1&0\\
-B&1\\
\end{array}
\right)\label{j12}
\ee
where $g$ is a K\"ahler metric, which is bihermitian with respect to both 
complex structures $J_\pm$, while $B$ is a 2-form field. We leave a discussion
about its relationship with the $B$-field of the sigma model for section 3.4.

First, we shall derive the conditions for a generic element of
 $(\xi,\pm id\mu)\in T\oplus T^*$ to be an eigenvector of
the generalized complex structures.
By identifying $\xi\in T$ with a Killing vector, we solve for the
one form $d\mu\in T^*$. Next, after verifying  that $d\mu$ is an exact
one-form, we shall compare it with the
the moment map and enquire whether these expressions are compatible.
We discuss two concrete settings: the almost product structure spaces, with
their commuting  complex structures, and as an example of
bihermitian geometry we turn to  the $SU(2)\times U(1)$ WZNW 
sigma model.

We begin with some formal statements.  The condition that an element
of $T\oplus T^*$ lies in the eigenbundle of  $\mcal{J}_1$ is
\be
\mcal{J}_1
\left(
\begin{array}{c}
\xi \\
icd\m \\
\end{array}
\right)
=
ai
\left(
\begin{array}{c}
\xi \\
icd\m \\
\end{array}
\right),
\ee
where $c=\pm1$, $a=\pm1$.
After a bit of massaging, we find that this eigenvalue problem is equivalent
to the following linear homogeneous equation system\footnote{For the eigenvalue
problem associated with the other generalized almost complex structure $
\mcal J_2$, we find a similar linear homogeneous system: $$(J_+-ai)(\G-\xi)=0,
\qquad (J_-+ai)(\G+\xi)=0.$$}
\bea
\label{eqts}
&& (J_+-ai)(\G-\xi)=0\cr
\cr
&& (J_--ai)(\G+\xi)=0,
\eea
where
\be
\label{M}
\G=G^{-1}(B\xi-icd\m)
\ee
Then, by solving (\ref{eqts}) we find $\xi$ and $\G$. The number of independent
solutions is equal to the number of zero eigenvalues of $J_\pm-ai$.
After identifying $\xi$ with a certain Killing vector,
we generically find a corresponding  $\G$. This allows us to solve for $\mu$:
\be
\label{form}
d\m=ic(G\G-B\x).
\ee
To test the compatibility between this expression and
the moment map (\ref{mommap}), in the next sections we explore two concrete
examples of bihermitian geometry.

\subsection{Specialization to spaces with almost product structure}
\indent

In the case of a space with almost product structure, which is realized by
a (2,2) sigma model written in terms of chiral and twisted chiral
superfields \cite{ghr}, we may choose to work
in a coordinate system where the two commuting complex structures are diagonal:
\bea
J_+&=&
\left(
\begin{array}{cc}
J_1&0\\
0&J_2\\
\end{array}
\right)~~~
J_-=
\left(
\begin{array}{cc}
J_1&0\\
0&-J_2\\
\end{array}
\right).
\eea
In the same coordinate system, the metric and $B$-field are also
block-diagonal:
\bea
g&=&
\left(
\begin{array}{cc}
g_1&0\\
0&g_2\\
\end{array}
\right)~~~
B=
\left(
\begin{array}{cc}
0&b\\
-b^t&0\\
\end{array}
\right).
\eea
The expressions taken by $g,B,J_+, \mbox{and } J_-$ suggest that we
should consider a similar decomposition for $\xi$,$\G$ and $d\m$. Specifically,
\be
\xi=
\left(
\begin{array}{c}
\xi_1\\
\xi_2\\
\end{array}
\right)~~~
\G=
\left(
\begin{array}{c}
\G_1\\
\G_2\\
\end{array}
\right)~~~
d\m=
\left(
\begin{array}{c}
d\m_1\\
d\m_2\\
\end{array}
\right).
\ee
Under this decomposition $\G_{1,2}, \xi_{1,2}$ are solutions to (\ref{eqts}):
\bea
\label{Vc}
&& (J_1-ai)\G_1=(J_1-ai)\xi_1=0\cr
&& ai\G_2=-J_2\xi_2.
\eea
and (\ref{form}) becomes
\bea
&& d\m_1=icg_1\G_1-icb\xi_2\cr
&& d\m_2=icg_2\G_2+icb^t\x_1,
\label{eigenvec}
\eea
How does this compare with the moment maps which are given by
$
d\mu_{\pm}=\o_{\pm}\xi \mp J_{\pm}^T u
$?
When we specialize to the case where the Lie derivative of $B$ with respect to $\xi$
vanishes, $\mcal{L}_{\xi} B=0$, we can use that $u=-B\xi+d\sigma$.
  Taking the appropriate linear
combinations that match up most closely with the generalized complex
structures we define 
\bea
&& d\tilde M=\frac{1}{2}(d\m_+ + d\m_-),\qquad
d\hat M=\frac{1}{2}(d\m_+ - d\m_-)
\eea
where
\bea
\label{mmc}
\left(
\begin{array}{c}
d\tilde M_1 \\
d\tilde M_2 \\
\end{array}
\right)
=
\left(
\begin{array}{c}
\o_1\xi_1 \\
-J^t_2b^t\xi_1 \\
\end{array}
\right),
\qquad
\left(
\begin{array}{c}
d\hat M_1 \\
d\hat M_2 \\
\end{array}
\right)
=
\left(
\begin{array}{c}
-J^t_1b\xi_2 \\
\o_2\xi_2 \\
\end{array}
\right).
\eea
The matching between (\ref{eigenvec}) and (\ref{mmc})  can be done using either $d\tilde{M}$ or 
$d\hat{M}$. For concreteness, we choose to match (\ref{eigenvec}) against $d\hat{M}$. This is done 
provided that $\G_1=\xi_1=0$. The condition $\xi_1=0$ is
automatically satisfied for almost product structure geometries, where
$J_{1,2}$ are both diagonal.
This is so because the requirement that $\xi$ is holomorphic (i.e. it leaves invariant
the complex structures) implies that that either
$\xi_1$ or $\xi_2$ vanish \cite{hps}.
Next to complete the matching of (\ref{mmc}) and (\ref{eigenvec}) we need
$\G_2=\pm i J_2\xi_2$, but is exactly the expression of $\G_2$
which we get from (\ref{Vc}).

Now that we have verified the compatibility of two moment map definitions,
 (\ref{form}) and (\ref{mommap}), for
the almost product structure geometry,
we want to investigate their compatibility in a more generic
case of bihermitian geometry.
Since the complex
structures do not commute in this case, it is difficult to analyze what
happens in general.  However we can consider the concrete $SU(2)\otimes U(1)$
example and see how things work out there.

\subsection{The $SU(2)\times U(1)$ example}
\indent

In this case, the non-commuting complex structures, read off from the
supersymmetry transformations of the non-linear sigma model
\cite{lruz,ikr}, are:
\bea
&&
J_+=
\left(
\begin{array}{cccc}
i& 0& 0& 0\\
0& -i& 0& 0\\
-2i& 0& -i& 0\\
0& 2i& 0& i\\
\end{array}
\right)
~J_-=
\left(
\begin{array}{cccc}
i& 0& 2i(1-f)& 0\\
0& -i& 0& -2i(1-f)\\
0& 0& -i& 0\\
0& 0& 0& i\\
\end{array}
\right),
\eea
where $f=f(\eta+\bar\eta)$.
The U(1) Killing vector is  $\xi=(i,-i,-i,i)$. The $B$-field was given in
(\ref{stuff}), and the metric takes the form
\be
g=\begin{pmatrix}
0&2&0&2(1-f)\\
2&0&2(1-f)&0\\0&2(1-f)&0&2(1-f)\\2(1-f)&0&2(1-f)&0\\
\end{pmatrix}.
\ee
The moment map $d\mu_+=\omega_+\xi -J_+^T u$ reads
\bea
d\mu_+&=&(-2f,-2f,0,0)-(-2f,-2f,-2f,-2f)-(iC_r-2iC_\eta,
-iC_{\bar r}+2iC_{\bar\eta},-iC_\eta,iC_{\bar\eta})\nonumber\\
&=&(iC_r-2iC_\eta,
-iC_{\bar r}+2iC_{\bar\eta},2f-iC_\eta,2f+iC_{\bar\eta}).
\label{mommapsu2}
\eea
wheare the last term on the first line
represents the ambiguity in $u$, $J_+^T d\sigma$. The constants $C_{r,\bar r,
\eta,\bar\eta}$ satisfy the constraint $C_r-C_{\bar r}-C_\eta+C_{\bar\eta}=0$.

We find that the solution to (\ref{eqts}), corresponding
to a $+i$ eigenvector, ($a=1$), is given by $(\xi, \G_{1,\pm})$,
where $\xi=(i,-i,-i,i)$ and
\be
\G_{1,+}=\left(
-i,-i,i,i\frac{1+f}{1-f}
\right).
\ee
For a $-i$ eigenvector ($a=1$), we find
\be
\G_{1,-}=\left(i,
i,
-i\frac{1+f}{1-f},
-i
\right),
\ee
for the same Killing vector $\xi$.
For completeness we record the eigenvectors $(\xi, \G_{2,\pm})$ of
the second generalized almost complex structure $\mcal J_2$:
$\G_{2,+}=(-i,-i,i,-i)$ corresponds to the $+i$ eigenvalue
and $\G_{2,-}=(i,i,i,-i)$ to the $-i$ eigenvalue.

{From} (\ref{mmc}), substituting $\G_{1,\pm}$ as well as the
the metric, $B$-field, and Killing vector we get
\be
ic G\G_{1,+}=c(-2f,2f,-4f,0), \qquad ic B\xi= c(-1+2f,1-2f,-1+2f,1-2f),\label{gvbx}
\ee
where we recall that $c=\pm1$. We have also identified the 2-form $B$
in the generalized almost complex structure with the $B$-field.
Notice that in order to
be able to recover an expression compatible with (\ref{mommapsu2}), we must
take  the {\it sum $ic(G\G+B\xi)$}, and not the difference of the two terms in
(\ref{gvbx})!  The reason for an apparent discrepancy between the two
expressions that we have for the moment map, (\ref{mommap}) and (\ref{form})
lies in the identification of the sigma model $B$ field and the 2-form
$B$ that appears in the generalized almost complex structure (\ref{j12}).
The agreement is restored upon {\it making the identification between
minus the sigma model $B$-field and the object by the same name present in
(\ref{j12}).} It is essential that in replacing $B\to -B$ in (\ref{j12}),
with $B$ the sigma model $B$-field, we haven't spoiled any of the properties
of the generalized K\"ahler geometry objects.

To complete our argument, we have to make the following assignments for the
constants which enter in the one-form $d\sigma$: $C_r=C_{\bar r}=C_\eta=
C_{\bar\eta}=i$.

In conclusion, we still find it possible to obtain the moment map from the
condition that together with the Killing vector forms a pair $(\xi, ic d\mu)$
which lies in the eigenbundle of the generalized almost complex structure.
However, we must exercise caution and  interpret the 2-form $B$ in
(\ref{j12}) as {\it minus} the sigma-model $B$-field. We have also seen that
the matching between (\ref{form}) and (\ref{mommap}) requires making use
of the ambiguity in defining the one-form $u$. The exact, U(1) invariant
one-form $d\sigma$ required by the matching between the two moment map
definitions led us to a different one-form $d\sigma$ than the one we
identified in Section  2.2 by matching $u$ with the gauged sigma model action.

\section{T Duality}
\indent

T-duality can be implemented, while preserving the manifest (2,2)
supersymmetries of the sigma model,
by performing a Legendre transformation of the K\"ahler
potential.
This procedure amounts to starting from the gauged sigma model,
introducing a Lagrange multiplier that enforces the condition that the
gauge field is pure gauge, and eliminating the gauge field from its
equation of motion. In terms of the geometric data, by descending
to the level of (1,1) superspace, we find that under T-duality, the
metric and B-field transform according to the Buscher rules.
Let us begin with some review material detailing the execution of T-duality
in (2,2) superspace.
The simplest example of T-duality involves a non-linear sigma model
written in terms of either chiral or
twisted chiral superfields with an U(1) isometry.
Under T-duality the chiral multiplets are mapped into twisted anti-chiral and
vice-versa.
Specifically, we choose a coordinate system such
that the isometry is realized by a shift in a particular coordinate.
Then the K\"{a}hler potential has the form
\be
\label{original}
K=K(\bP+\P,Z^a),
\ee
where $Z^a$ are spectator fields that can be either chiral or twisted chiral.
According to the discusion in Section 2.3, the gauged action is obtained
 by replacing $\bP+\P$ with $\bP+\P+V$ where $V$ is the
usual superfield prepotential for the gauge multiplet. The
gauged K\"ahler potential is
\be
K_g=K(\bP+\P+V,Z^a).
\ee
To construct the duality functional we introduce a Lagrange multiplier
that forces the gauge multiplet field strength to vanish:
\be
K_D=K_g+U(S+iP)+\bar U(S-iP).
\ee
Since $(S+iP)=\frac{i}{2}\bar D_+D_-V$ we see that the $U$ and
$\bar U$ equations of motion force $V$ to be pure gauge, i.e., $V=\L+\bar\L$,
with $\Lambda$ a chiral superfield.  For the next step, by
 choosing a gauge such that $\Phi+\bar\Phi$ have been completely gauged away
\be
K_{g}=K(V,Z^a)
\ee
we arrive at the duality functional
\be
K_D= K(V,Z^a)-U(S+iP)-\bar U(S-iP).\label{dualfunc}
\ee
The original K\"ahler potential is recovered by integrating out $U$ and
$\bar U$.
The T-dual theory is obtained by integrating out the gauge field.
Its equation of motion is
\be
\frac{\partial K}{\partial V}-(\Psi+\bar\Psi)=0,
\ee
where $\Psi=\frac{i}{2}\bar D_+D_-U$ is a twisted anti-chiral superfield.
This defines $V=V(\Psi+\bar\Psi,Z^a)$.  The dual potential
\be
\tilde K=K(V,Z^a)-(\Psi+\bar\Psi)V
\ee
is the Legendre transform of the original potential (\ref{original}).

When one introduces semi-chiral superfields the story becomes somewhat more
complicated.
In \cite{gmst}, Grisaru et al. gave a detailed discussion of the various
descriptions of a (2,2) sigma model, which can be obtained by means of a
Legendre transform. Starting with a (2,2) K\"ahler potential written in terms
of semi-chiral superfields $K(X,\bar X,Y,\bar Y)$,  one constructs
the duality functional
\be
K(r,\bar r,s,\bar s)-Xr-\bar X\bar r-sY-\bar s\bar Y\label{grisaru}
\ee
where $r,\bar r, s ,\bar s$ are unconstrained superfields. Depending
which fields are integrated out $(X,Y),(r,s),(r,Y),(s,X)$ one finds
four equivalent formulations. In the absence of isometries, this amounts
to performing a sigma-model coordinate transformation.
The authors of \cite{gmst} investigated the consequences that the existence of
an isometry have on the duality functional.
For instance if the K\"ahler potential has a U(1) isometry
$K=K(X+\bar X,X+\bar Y,\bar X+Y)$, the duality functional
reads $K(r+\bar r, \bar r+s,r+\bar s)-(X+\bar X-Y-\bar Y)(r+\bar r)/2+
(X-\bar X+Y-\bar Y)(r-\bar r)/2-(r+\bar s)Y-(\bar r +s)\bar Y$.
By integrating over $r-\bar r$,
ultimately leads to expressing $X$ and $Y$ as the sum and difference
of a chiral and twisted chiral superfield. In this case,
the dual description of the
sigma model involves chiral and twisted chiral superfields. The $SU(2)\times
U(1)$ WZNW model has two such dual descriptions \cite{ikr}. The geometry
does not change as we pass from one description to the other, but
the pair of complex structures does change, from non-commuting complex
structures, to commuting ones.

On the other hand, not all the dualities following from
(\ref{grisaru}) can be derived from gauging an isometry.
The reason is that Lagrange multipliers in (\ref{grisaru}) are
semi-chiral superfields.
Following the discussion given at the beginning of this section,
one would need a gauge multiplet with a semi-chiral field strength,
in order to cast the gauged action duality functional (\ref{dualfunc})
into (\ref{grisaru}).
However, no known $(2,2)$ gauge multiplet contains such a
field strength.

Therefore we choose to pursue the construction of the
T-dual action of a sigma model with semi-chiral multiplets
following the steps which led to (\ref{dualfunc}).
We add Lagrange multiplier terms to the gauged action as
described previously, and construct the duality functional
as in \cite{rv}.
However, a technical difficulty, related to gauge fixing,
 prevents a straightforward
application of this procedure. Let us explain.

The U(1) invariant K\"ahler
potential, which generically takes the form given in (\ref{isok}),
can be gauged by adding the prepotential $V$ to
the appropriate field combinations.  The gauged K\"ahler potential is
$
K_g=K(X+\bar X+V, Y+\bar Y+V, X+Y+V).
$
Because the semi-chiral superfield is not
generically reducible in terms of chiral and twisted chiral
superfields\footnote{We thank Martin Ro\v{c}ek for explaining this
point to us.} one cannot completely gauge away $X$ or $Y$, as it was possible
for the chiral and/or twisted chiral superfields.
Trying to gauge away $X$ we could fix $X|=D_\a X|=D^2X|=0$, where
$|$ means evaluation with all the Grassmann
variables set to zero.  Since $X$ has
higher order components which are independent of the lower components we
realized that we have not gauged away all the $X$ components.  The independent
left over components form a $(1,1)$ Weyl spinor superfield.
We shall address the resolution to this question in the following section.

\subsection{Dualizing With Chiral and Twisted Chiral Superfields.}
\indent

For simplicity we will consider a K\"{a}hler potential,
parameterized by chiral and twisted chiral superfields,
which is  strictly invariant under the isometry.     The
potential is given by (\ref{original}). We begin in the slightly more
general setting:
\be
K_g=K(\bP+\P,Z^a)+\frac{e^L-1}{L}VM.
\ee
The moment map, $M$, is given by $M=i\xi_{\bar c}K$, and in this case
$\xi_{\bar c}=-i\frac{\partial}{\partial\bP}$.  To construct the duality
functional we add Lagrange multiplier terms that force the superfield
strength to vanish.  This gives the Lagrangian
\be
K_D=K(\bP+\P,Z^a)+\frac{e^L-1}{L}VM+(\bar\Psi+\Psi)V.
\ee
The final step is chosing a gauge.  Instead of setting $\P+\bar\P=0$,
we choose the Wess-Zumino gauge for the prepotential
$V$
\be
V|=D_\a V|=D^2V|=0.
\ee
This gauge choice will allow a better comparison with the semi-chiral case.
To see that we do get back the original Lagrangian, we
integrate out $\Psi$ and $\bar\Psi$. This implies that
\be
V=\bar\L+\L,
\ee
where $\L$ is a chiral superfield.  However, consistency with the gauge
choice requires that $V=0$ and this give us back the original K\"{a}hler
potential.  To find the dual potential we integrate out $V$.  Since
$(V)^3=0$ in the Wess-Zumino gauge, this allows
us to solve for $V$ explicity.  We obtain
\bea
&&V=i\frac{\bar\Psi+\Psi+M}{\xi_{\bar c}M}\cr
&&\tilde K=K(\bP+\P,Z^a)+\frac{i}{2}\frac{(\bar\Psi+\Psi+M)^2}{\xi_{\bar c}M}.
\eea
The important thing to note here is that consistency of the solution for V
with the gauge fixing conditions require that
\be
V|=0=i\frac{\bar\Psi+\Psi+M}{\xi_{\bar c}M}|\Ra (\bar\Psi+\Psi)|=-M|
\ee
It should be understood that this is a component equation, and
not a superfield equation.  With this in hand we can show the following;
\bea
\label{cond}
\frac{\partial^2\tilde K}{\partial\bP\partial\P}|=0,\qquad
\frac{\partial^2\tilde K}{\partial Z^a\partial\P}|=0,\qquad
\frac{\partial^2\tilde K}{\partial\bar\Psi\partial\P}|=-1.
\eea
The implication which follows from these equations is that the contribution
of $\P |$ to the geometry has been
replaced by $\Psi |$ up to a surface term that comes from the new $B$ field.
Let us demonstrate how this works with a simple example,
specifically $R\ra\frac{1}{R}$
for one of the cycles on $T^2$.  The K\"{a}hler potential and moment map are:
\bea
&&K=\frac{R}{2}(\bP+\P)^2\cr
\cr
&&M=R(\bP+\P).
\eea
The dual potential is
\be
\tilde K=-\frac{1}{2R}(\bar\Psi+\Psi)^2-(\bP+\P)(\bar\Psi+\Psi).
\ee
While this looks as though both directions of $T^2$ were dualized,
one must remember that
the real part of $\Psi |$ is proportional to $R$ times the real part
of $\P |$.  Only the direction parameterized by the imaginary part of $\P |$
was dualized.
\subsection{Dualizing with semi-chiral superfields}
\indent

Now we can give a straightforward extension of the previous discussion
to the case when we dualize an isometry of a sigma
model parametrized by semi-chiral superfields.
We start with equation (\ref{geda}), add the Lagrange multipliers enforcing
that $V$ is pure gauge, and choose the same gauge Wess-Zumino gauge as in the
previous section.  The dual potential Kahler potential is:
\be
\tilde K=K(X,\bar X,Y,\bar Y,Z^a)+\frac{i}{2}\frac{(\bar\Psi+\Psi+M)^2}
{\x_{\bar c}M}.
\label{tdualKahler}
\ee
The analogue of (\ref{cond}) reads:
\bea
(\x_c)(\x_{\bar c})\tilde K|=0,\qquad
\frac{\partial (i\x_{\bar c}\tilde K)}{\partial Z^a}|=0\qquad
\frac{\partial (i\x_{\bar c}\tilde K)}{\partial \Psi}|=-1.\label{tsc}
\eea
{From} (\ref{tsc}) we see that the coordinates in the combination of
semi-chiral
superfields corresponding to $\x_c$ have been replaced by coordinates in a
twisted chiral superfield in the dual geometry.  This is analogous to what
happened in the case of  chiral and twisted chiral superfields. It was also
expected from gauge fixing considerations, although it was not a propri
clear exactly how it would happen.
We now have an explicit description of the T dual of a
theory with semi-chiral superfields at the manifest $(2,2)$ sigma model level.
\subsection{An example: T-duality with semi-chiral superfields in flat space}

\indent

In this section we try to develop some intuition about the dualization
prescription described in the previous section. Given that we perform a
duality transformation by gauging away {\it part} of a certain combination
of semi-chiral superfields, and in doing so we trade it for a
twisted chiral superfield, it is not
a priori obvious that this is equivalent to the Buscher rules.
In particular, we would like to check this in a simple example, namely
flat space with a U(1) isometry.

We start with four-dimensional flat space as our simplest example
because one needs both left and right pairs of chiral and anti-chiral superfields
in order to be able to eliminate the auxiliary components of the semi-chiral
superfields and obtain a sigma-model action.  Therefore we begin with
the following (2,2) K\"{a}hler potential
\be
K=R(\bar X+\bar Y)(X+Y)-\frac{R}{4}(Y+\bar Y)^2
\ee
where $\bar D_+ X=D_-Y=0$.  By descending to the level of (1,1) superspace
using \cite{lruz2}, we find the sigma model metric
\be
G=\left(
\begin{array}{cccc}
0 & 2R & R & 0\\
2R & 0 & 0 & R\\
R & 0 & 0 & R\\
0 & R & R & 0\\
\end{array}
\right),
\ee
where the rows and columns are labelled by $X|,\bar X|, \bar Y|, Y|$.
This gives us the action for the bosonic components
\be
S=\int d^2\s R(\del^aX\del_a\bar X+\del^a(\bar X+\bar Y)\del_a(X+Y)),
\ee
where for simplicity we denoted by $X$ the bosonic component of the
(1,1) superfield $X|$.
Denoting $Z=X+Y$ we notice that it is inert under the global shift symmetry.
By performing a diffeomorphism transformation to $(X,\bar X, Z, \bar Z)$,
we obtain the metric in canonical form
\be
G=\left(
\begin{array}{cccc}
0 & R & 0 & 0\\
R & 0 & 0 & 0\\
0 & 0 & 0 & R\\
0 & 0 & R & 0\\
\end{array}
\right).
\ee
The T-dual sigma model is obtained from the dual (2,2) Kahler potential
given in (\ref{tdualKahler}). In this particular case, (\ref{tdualKahler}) reads:
\be
\tilde K=R(\bar X+\bar Y)(X+Y)-\frac{R}{4}(Y+\bar Y)^2-\frac{1}{3R}
\bigg(\psi+\bar\psi+R(\bar X+X+\frac{1}{2}(\bar Y+Y))\bigg)^2
\ee
and the corresponding T-dual sigma-model metric is equal to:
\be
\frac{9}{2}G=\left(
\begin{array}{cccccc}
 -4 R & 5 R & 4 R & -5 R & 5 & -4 \\
 5 R & -4 R & -5 R & 4 R & -4 & 5 \\
 4 R & -5 R & -4 R & 5 R & -5 & 4 \\
 -5 R & 4 R & 5 R & -4 R & 4 & -5 \\
 5 & -4 & -5 & 4 & -\frac{4}{R} & \frac{14}{R} \\
 -4 & 5 & 4 & -5 & \frac{14}{R} & -\frac{4}{R}
\end{array}
\right),
\ee
where the rows and columns are labelled by $X,\bar X,\bar Y, Y,\psi,
\bar\psi$.
At first sight this result is puzzling, because we claim that we found the
T-dual of a  sigma model whose target space is flat four-dimensional space.
At the same time, the dual sigma-model involves six fields, and so,
apparently the target space is six-dimensional. These two seemingly
contradictory statements are reconciled when one takes a closer look at the
T-dual metric, and finds that it actually describes a four dimensional
subspace. This is obvious when expressing the previous T-dual metric
in terms of the following coordinates: $(X,\bar X, W=Y-\bar X, \bar W,
\psi,\bar\psi)$
\be
\frac{9}{2}G=\left(
\begin{array}{cccccc}
 0 & 0 & 0 & 0 & 0 & 0 \\
 0 & 0 & 0 & 0 & 0 & 0 \\
 0 & 0 & -4 R & 5 R & 4 & -5 \\
 0 & 0 & 5 R & -4 R & -5 & 4 \\
 0 & 0 & 4 & -5 & -\frac{4}{R} & \frac{14}{R} \\
 0 & 0 & -5 & 4 & \frac{14}{R} & -\frac{4}{R}
\end{array}
\right),
\ee
where we make the observation that $W=Y-\bar X$ is also inert under the global
U(1) action. The final step in getting the metric in its canonical form is
to make a coordinate transformation to $T=W-\frac{1}{R}\psi$:
\be
\frac{9}{2}G=\left(
\begin{array}{cccccc}
 0 & 0 & 0 & 0 & 0 & 0 \\
 0 & 0 & 0 & 0 & 0 & 0 \\
 0 & 0 & -4 R & 5 R & 0 & 0 \\
 0 & 0 & 5 R & -4 R & 0 & 0 \\
 0 & 0 & 0 & 0 & 0 & \frac{9}{R} \\
 0 & 0 & 0 & 0 & \frac{9}{R} & 0
\end{array}
\right).
\ee
This form of the T-dual metric makes it clear that the T-dual geometry
is four-dimensional and that the Buscher rules, which in this case amount to
$R\to 1/R$, are obeyed.

The reason why extracting the T-dual geometry required some work on our part
is that the semi-chiral superfields give rise to two (1,1) superfields,
one of them being auxiliary. As we gauge an isometry with (2,2) vector
superfields, the Lagrange multipliers enforcing the condition that the
vector field is pure gauge are
twisted chiral superfields (or chiral superfields). To some extent, as we
dualize, we exchange the coordinates in a combination of semi-chiral
superfields by coordinates in a twisted chiral superfields. However, due
to the presence of the auxiliary superfields, even the coordinates which are
not directly affected by the duality (like the $Z,\bar Z$ coordinates)
end up mixing with the dualized coordinate.

\section{Conclusions}
\indent

In this paper we have continued the ongoing investigation of the
connection between generalized K\"{a}hler
geometry and two-dimensional $\mcal{N}=(2,2)$ non-linear sigma models.
Specifically, we addressed aspects in the area concerning
gauged sigma models with semi-chiral superfields and its relation to
various geometric structures on the mathematical side.
We have given the form of
the gauged action
in $(1,1)$ and $(2,2)$ superspace and identified the moment map  as well as the one-form needed 
for the gauging the sigma model, as demonstrated by Hull and Spence in \cite{hs}.  In two particular
cases, namely spaces with almost product structure and the
$SU(2)\times U(1)$ WZNW model written in terms of semi-chiral superfields,
we have found that the combination
of the moment map and Killing vector associated to the isometry $\xi -\sqrt{-1}d\mu^\xi$ does lie in the eigenbundle of
the generalized almost complex structures, as stipulated by the definitions in the mathematical literature. 
We leave for a future publication the
relationship between the physical moment map and the mathematical definition (in the general case).
Finally, we have
presented a description
of T-duality for generic $(2,2)$ sigma models with manifest $(2,2)$
supersymmetry.
It is interesting
to note that T-duality, as we have described it, introduces a twisted
chiral superfield into the sigma model.
We could also have described it in such a way that a chiral
superfield would be introduced.
It would also be interesting to explore the  changes which
the sigma model geometry undergoes since it appears that the superfield representations for the coordinates get mixed
due to T-duality.

\section*{Acknowledgments}

We are grateful to  L. Anguelova, D. Belov, S.J. Gates,  V. Mathai, M. Ro\v{c}ek 
and B. Uribe for comments and suggestions.
W.M. thanks the MCTP for hospitality during different stages of
this work. WM and LAPZ also thank
the KITP for hospitality during the early stages of this work while both
were participants of the program
``Mathematical Structures in String Theory''.
This work is  partially supported by Department of Energy under
grant DE-FG02-95ER40899 to the University of Michigan and by the
National Science Foundation under rant No. PHY99-07949.


\begin{thebibliography}{99}


\bibitem{zumino}
  B.~Zumino,
 ``Supersymmetry And Kahler Manifolds,''
  Phys.\ Lett.\ B {\bf 87} (1979) 203.


\bibitem{ag-f}
  L.~Alvarez-Gaume and D.~Z.~Freedman,
 ``Ricci Flat Kahler Manifolds And Supersymmetry,''
  Phys.\ Lett.\ B {\bf 94} (1980) 171.

\bibitem{lr}
  U.~Lindstrom and M.~Rocek,
  ``Scalar Tensor Duality And N=1, N=2 Nonlinear Sigma Models,''
  Nucl.\ Phys.\ B {\bf 222}, 285 (1983).
  
  \bibitem{rt}
  M.~Rocek and P.~K.~Townsend,
  Phys.\ Lett.\ B {\bf 96}, 72 (1980).

 \bibitem{cf}
  T.~L.~Curtright and D.~Z.~Freedman,
  Phys.\ Lett.\ B {\bf 90}, 71 (1980)
  [Erratum-ibid.\ B {\bf 91}, 487 (1980)].
  
  \bibitem{hypereview}
  N.~J.~Hitchin, A.~Karlhede, U.~Lindstrom and M.~Rocek,
``Hyperkahler Metrics And Supersymmetry,''
  Commun.\ Math.\ Phys.\  {\bf 108} (1987) 535.
  
\bibitem{ag-f-2}
  L.~Alvarez-Gaume and D.~Z.~Freedman,
   ``Geometrical Structure And Ultraviolet Finiteness In The Supersymmetric
  Sigma Model,''
  Commun.\ Math.\ Phys.\  {\bf 80}, 443 (1981).


\bibitem{ghr}
  S.~J.~.~Gates, C.~M.~Hull and M.~Rocek,
 ``Twisted Multiplets And New Supersymmetric Nonlinear Sigma Models,''
  Nucl.\ Phys.\ B {\bf 248} (1984) 157.

\bibitem{blr}
  T.~Buscher, U.~Lindstrom and M.~Rocek,
 ``New Supersymmetric Sigma Models With Wess-Zumino Terms,''
  Phys.\ Lett.\ B {\bf 202} (1988) 94.

\bibitem{hit}
  N.~Hitchin,
  ``Generalized Calabi-Yau manifolds,''
  Quart.\ J.\ Math.\ Oxford Ser.\  {\bf 54}, 281 (2003)
  [arXiv:math.dg/0209099].
  
 \bibitem{thesis}
M. Gualtieri, "Generalized Complex Geometry", math.DG/0401221.

\bibitem{bursztyn-crainic}
H.Bursztyn and M. Crainic, ``Dirac structures, moment maps and quasi-Poisson manifolds,''
 math.DG/0310445

\bibitem{crainic}
 M. Crainic, ``Generalized complex structures and Lie brackets,'' math.DG/0412097

\bibitem{lin-tolman}
Y. Lin and  S. Tolman, ``Symmetries in generalized K\"{a}hler geometry,'' math.DG/0509069


\bibitem{lin-tolman2}
Y. Lin and S. Tolman, ``Reduction of twisted generalized K\"ahler structure,'' math.DG/0510010


\bibitem{hu}
S. Hu, ``Hamiltonian symmetries and reduction in generalized geometry,'' math.DG/0509060.

\bibitem{bursztyn-cavalcanti-gualtieri}
H. Bursztyn, G. Cavalcanti and  M. Gualtieri, ``Reduction of Courant algebroids and generalized complex structures,''
math.DG/0509640

\bibitem{lmtz}
  U.~Lindstrom, R.~Minasian, A.~Tomasiello and M.~Zabzine,
  ``Generalized complex manifolds and supersymmetry,''
  Commun.\ Math.\ Phys.\  {\bf 257}, 235 (2005)
  [arXiv:hep-th/0405085].
  
  \bibitem{lruz}
  U.~Lindstrom, M.~Rocek, R.~von Unge and M.~Zabzine,
``Generalized Kaehler geometry and manifest N = (2,2) supersymmetric
nonlinear sigma-models,''
  JHEP {\bf 0507} (2005) 067
  [arXiv:hep-th/0411186].

\bibitem{lruz2}
  U.~Lindstrom, M.~Rocek, R.~von Unge and M.~Zabzine,
  ``Generalized Kaehler manifolds and off-shell supersymmetry,''
  arXiv:hep-th/0512164.
  
  \bibitem{blp}
  A.~Bredthauer, U.~Lindstrom and J.~Persson,
  ``First-order supersymmetric sigma models and target space geometry,''
  JHEP {\bf 0601}, 144 (2006)
  [arXiv:hep-th/0508228].

\bibitem{blpz}
  A.~Bredthauer, U.~Lindstrom, J.~Persson and M.~Zabzine,
  ``Generalized Kaehler geometry from supersymmetric sigma models,''
  arXiv:hep-th/0603130.
  
  \bibitem{bhc2}
  N.~Hitchin,
 ``Instantons, Poisson structures and generalized Kaehler geometry,''
  arXiv:math.dg/0503432.

\bibitem{hps}
  C.~M.~Hull, G.~Papadopoulos and B.~J.~Spence,
 ``Gauge symmetries for (p,q) supersymmetric sigma models,''
  Nucl.\ Phys.\ B {\bf 363}, 593 (1991).

\bibitem{hklr}
  C.~M.~Hull, A.~Karlhede, U.~Lindstrom and M.~Rocek,
  ``Nonlinear Sigma Models And Their Gauging In And Out Of Superspace,''
  Nucl.\ Phys.\ B {\bf 266}, 1 (1986).
    
\bibitem{bw}
  J.~Bagger and E.~Witten,
  ``The Gauge Invariant Supersymmetric Nonlinear Sigma Model,''
  Phys.\ Lett.\ B {\bf 118}, 103 (1982).
    
 \bibitem{hs}
  C.~M.~Hull and B.~J.~Spence,
  ``The gauged nonlinear sigma model with Wess-Zumino term,''
  Phys.\ Lett.\ B {\bf 232}, 204 (1989).
   
\bibitem{rv}
  M.~Rocek and E.~P.~Verlinde,
``Duality, quotients, and currents,''
  Nucl.\ Phys.\ B {\bf 373}, 630 (1992)
  [arXiv:hep-th/9110053].

\bibitem{hs1}
  C.~M.~Hull and B.~J.~Spence,
  ``The Geometry of the gauged sigma model with Wess-Zumino term,''
  Nucl.\ Phys.\ B {\bf 353}, 379 (1991).

\bibitem{gates}
  S.~J.~J.~Gates,
   ``Vector multiplets and the phases of N=2 theories in 2-D: Through the
  looking glass,''
  Phys.\ Lett.\ B {\bf 352}, 43 (1995)
  [arXiv:hep-th/9412222].

\bibitem{agf}
  L.~Alvarez-Gaume and D.~Z.~Freedman,
  ``Potentials For The Supersymmetric Nonlinear Sigma Model,''
  Commun.\ Math.\ Phys.\  {\bf 91}, 87 (1983).

\bibitem{Jourjine:1983xs}
  A.~N.~Jourjine,
   ``The Effective Potential In Extended Supersymmetric Nonlinear Sigma
  Annals Phys.\  {\bf 157}, 489 (1984).
 
\bibitem{ikr}
  I.~T.~Ivanov, B.~b.~Kim and M.~Rocek,
``Complex structures, duality and WZW models in extended superspace,''
  Phys.\ Lett.\ B {\bf 343}, 133 (1995)
  [arXiv:hep-th/9406063].

\bibitem{st}
  A.~Sevrin and J.~Troost,
``Off-shell formulation of N = 2 non-linear sigma-models,''
  Nucl.\ Phys.\ B {\bf 492} (1997) 623
  [arXiv:hep-th/9610102].\\
  A.~Sevrin and J.~Troost,
 ``The geometry of supersymmetric sigma-models,''
  arXiv:hep-th/9610103.

\bibitem{gmst}
  M.~T.~Grisaru, M.~Massar, A.~Sevrin and J.~Troost,
  ``Some aspects of N = (2,2), D = 2 supersymmetry,''
  Fortsch.\ Phys.\  {\bf 47}, 301 (1999)
  [arXiv:hep-th/9801080].

\bibitem{hu2}
S. Hu, ``Reduction and duality in generalized geometry,'' math.DG/0512634

















\end{thebibliography}
\end{document}